\PassOptionsToPackage{table}{xcolor}
\documentclass[twocolumn]{aastex63}

\usepackage[table]{xcolor}
\usepackage{amsmath}
\newcommand{\dd}[1]{\mathrm{d}#1}

\makeatletter
\providecommand\add@text{}
\newcommand\tagaddtext[1]{%
  \gdef\add@text{#1\gdef\add@text{}}}%
\renewcommand\tagform@[1]{%
  \maketag@@@{\llap{\add@text\quad}(\ignorespaces#1\unskip\@@italiccorr)}%
}
\makeatother

\submitjournal{ApJ}
\accepted{November 11, 2020}
\shortauthors{Duncan et al.}

\begin{document}

\title{NuSTAR Observation of Energy Release in Eleven Solar Microflares}

\author[0000-0002-6872-4406]{Jessie Duncan}
\affiliation{University of Minnesota, Minneapolis, MN, USA}

\author[0000-0001-7092-2703]{Lindsay Glesener}
\affiliation{University of Minnesota, Minneapolis, MN, USA}

\author[0000-0002-1984-2932]{Brian W. Grefenstette}
\affiliation{California Institute of Technology, Pasadena, CA, USA}

\author[0000-0002-7407-6740]{Juliana Vievering}
\affiliation{University of Minnesota, Minneapolis, MN, USA}

\author[0000-0003-1193-8603]{Iain G. Hannah}
\affiliation{University of Glasgow, Glasgow, UK}

\author[0000-0002-0542-5759]{David M. Smith}
\affiliation{University of California at Santa Cruz, Santa Cruz, CA, USA}

\author{ S{\"a}m Krucker}
\affiliation{University of California at Berkeley, Berkeley, CA, USA}
\affiliation{University of Applied Sciences and Arts Northwestern Switzerland, Windisch, Switzerland}

\author[0000-0002-8574-8629]{Stephen M. White}
\affiliation{Air Force Research Laboratory, Albuquerque, NM, USA}

\author[0000-0001-5685-1283]{Hugh Hudson}
\affiliation{University of California at Berkeley, Berkeley, CA, USA}
\affiliation{University of Glasgow, Glasgow, UK}



\begin{abstract}

Solar flares are explosive releases of magnetic energy. Hard X-ray (HXR) flare emission originates from both hot (millions of Kelvin) plasma and nonthermal accelerated particles, giving insight into flare energy release. The \textit{Nuclear Spectroscopic Telescope ARray} (\textit{NuSTAR}) utilizes direct focusing optics to attain much higher sensitivity in the HXR range than that of previous indirect imagers. This paper presents eleven \textit{NuSTAR} microflares from two active regions (AR 12671 on 2017 August 21, and AR 12712 on 2018 May 29). The temporal, spatial, and energetic properties of each are discussed in context with previously published HXR brightenings. They are seen to display several `large-flare' properties, such as impulsive time profiles and earlier peaktimes in higher energy HXRs. For two events where active region background could be removed, microflare emission did not display spatial complexity: differing \textit{NuSTAR} energy ranges had equivalent emission centroids. Finally, spectral fitting showed a high energy excess over a single thermal model in all events. This excess was consistent with additional higher-temperature plasma volumes in 10/11 microflares, and consistent only with an accelerated particle distribution in the last. Previous \textit{NuSTAR} studies focused on one or a few microflares at a time, making this the first to collectively examine a sizable number of events. Additionally, this paper introduces an observed variation in the \textit{NuSTAR} gain unique to the extremely low-livetime (\textless 1\%) regime, and establishes a correction method to be used in future \textit{NuSTAR} solar spectral analysis.

\end{abstract}



\section{Introduction} \label{sec:intro}

Solar flares are dramatic manifestations of change in the magnetic structure of the solar corona. They have been observed across over seven orders of magnitude in estimated \textit{GOES (Geostationary Operational Environmental Satellite)} soft X-ray (SXR) flux. The accepted model of flare production involves energy released by magnetic reconnection \citep[e.g.][]{benzFO}. During this process, particles are accelerated to high energies by dynamic fields and emit bremsstrahlung radiation through interactions with ambient coronal plasma \citep[e.g.][]{Brown_1971}. \par

In addition to this nonthermal emission, flares also show significant thermal emission from plasma heated to high temperatures as a result of various mechanisms of energy release. Both nonthermal emission and thermal emission from the hottest flare plasma (millions of Kelvin) are evident in the hard X-ray (HXR) band, with nonthermal emission dominating at the highest energies \citep{emslie}. \par
\begin{figure*}
\centering
\includegraphics[width=\textwidth, scale=1]{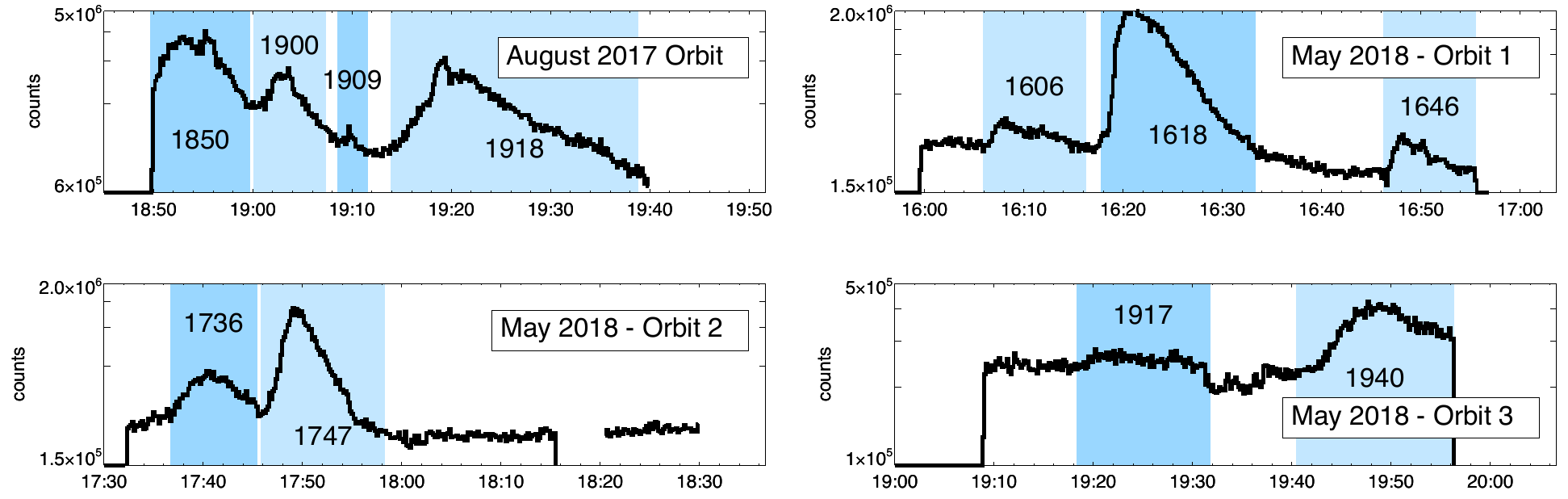}
\vspace{-0cm}
\caption{\textit{NuSTAR} emission between 2-10 keV is shown for the four orbits containing the 11 microflares. UTC times are shown in HH:MM format. Each lightcurve has been livetime-corrected, and binned in 10s intervals. Microflaring intervals are marked in blue, and labeled for future reference (these short flare IDs are adopted for reference throughout this work; refer to Table \ref{models} for standard Solar Object Locator target IDs). Differing y-scale between each of the four full-orbit lightcurves is noted, reflecting variable levels of activity in AR 12712 at different times (among the May 2018 orbits), and the comparatively higher level of activity observed from AR 12671 in August 2017.}
\label{fig:orbits}
\vspace{-0cm}
\end{figure*}

The \textit{Reuven Ramaty High-Energy Solar Spectroscopic Imager (RHESSI)} operated from 2002 to 2018, and allowed for extensive investigation of large flares using an indirect Fourier imaging method to observe from 3 keV to 17 MeV \citep{2002SoPh..210....3L}. In addition, \textit{RHESSI} was used for statistical HXR studies of \textit{GOES} A- and B-class microflares \citep{2008ApJ...677.1385C, Hannah_2008, 2011SSRv..159..263H}. However, the large detector volume required by \textit{RHESSI}'s imaging method caused a high background that limited the instrument's sensitivity to fainter events. \par

In recent years, the \textit{Nuclear Spectroscopic Telescope ARray} (\textit{NuSTAR}) satellite mission and the \textit{Focusing Optics X-ray Solar Imager (FOXSI)} rocket campaigns have demonstrated the significantly greater sensitivity possible with direct focusing HXR optics \citep{2013ApJ...770..103H, foxsi2016}. These instruments allow unprecedented opportunity for analysis of microflares, events with energy content estimated to be around six orders of magnitude less than that of the brightest solar flares. \textit{NuSTAR} and \textit{FOXSI} are capable of observing the very faintest A-class events, as well as brightenings that are too faint to be observed by \textit{GOES}.
\par

Flare occurrence rate is inversely related to magnitude, with fainter events observed far more frequently than brighter ones \citep[e.g.][]{2011SSRv..159..263H}. Because of this, a large ensemble of extremely faint flaring events, or nanoflares, are a theorized mechanism for observed large-scale heating of the solar corona \citep[e.g.][]{1988ApJ...330..474P, Klimchuk_2006}. Nanoflares would be faint and frequent enough that it would currently be impossible to detect them individually. They would occur across the entire corona, even in quiet regions with little evidence of large-scale magnetic activity. In this way, their combined effect could provide the energy necessary for coronal heating, which is not sufficiently accounted for by the energy released in observed flare populations \citep[e.g.][]{1991SoPh..133..357H}. \par

It has been proposed that nanoflares originate from a reconnection process similar to that of standard flares, but at a much smaller energy scale \citep[e.g.][]{1988ApJ...330..474P}. It remains unknown how frequently they might occur, whether they also accelerate particles, and the amount of energy that they could release. To refine our understanding of the emission we might expect from nanoflares, it is essential to investigate how flare properties change over a wide range of magnitudes. This particularly motivates the study of small microflares observed by current-generation focusing instruments, the faintest events ever observed in HXRs. \par

This paper provides detailed analysis of eleven microflares observed by \textit{NuSTAR}, with emphasis on characterization of their higher-energy spectral properties and examination of the correspondence between their temporal and spatial properties and those of larger flares. To provide context, Section \ref{nustarsolar} will present an overview of the process and history of \textit{NuSTAR} solar observation, and also introduce the host of microflares. Sections \ref{time} and \ref{space} will include consideration of their temporal and spatial properties in context with those of larger `standard' size flares. Finally, Section \ref{spec} will consider the spectral properties of each observed event, determining for each whether the emission is best explained by a multi-thermal plasma model alone, or by a combination of thermal and nonthermal components. \par


\section{\textit{NuSTAR} Solar Observation} \label{nustarsolar}

\par \textit{NuSTAR} is a NASA Small Explorer mission launched in 2012. It has two co-aligned focusing X-ray optics designed to observe in the 3-79 keV band (with the range down to 2.5 keV usable in some high-flux observations \citep{Grefenstette_2016}), 18$"$ angular resolution (FWHM), and a 12$'$x12$'$ field of view (FOV) \citep{2013ApJ...770..103H}. Data from the two telescopes are identified by reference to the focal plane module (FPM) associated with each detector (FPMA, FPMB). \textit{NuSTAR} is an astrophysical mission, and as such faces limitations when used for solar observation \citep{Grefenstette_2016}. In particular, high flux rates can cause low detector livetime when observing brighter solar events, making \textit{NuSTAR} primarily suitable for observation of small flares, quiescent active regions, and the quiet Sun. \textit{NuSTAR} also has a pointing uncertainty of up to a few arcminutes in absolute astrometry when observing the Sun, as its forward-facing star-tracking camera is blinded by the solar disk \citep[e.g.][]{Glesener_2017}. To mitigate this uncertainty, \textit{NuSTAR} data can be co-aligned with extreme ultraviolet (EUV) context data. \par

\textit{NuSTAR} can experience abrupt jumps in pointing associated with changes in the combination of star-tracking camera head units (CHUs) being used to determine its orientation \citep{Grefenstette_2016}. The occurrence of these shifts is well documented within \textit{NuSTAR} data structures, and is considered in every stage of the analysis process. CHU shifts can restrict which time intervals can be easily used for spectroscopy. \par

Despite these limitations, \textit{NuSTAR} has completed a growing number of solar observation campaigns over the last few years, many of which have included observation of both active region microflares and quiet Sun brightenings. The magnitudes of these small events are generally compared in terms of their \textit{GOES} class, a flare classification scheme based on X-ray brightness in the 1-8 \AA\  range as observed by \textit{GOES} satellites. \textit{NuSTAR} microflares studied so far have all been A-class or smaller, implying a brightness below 10\textsuperscript{-7} watts m\textsuperscript{-2} (A-class events), or below 10\textsuperscript{-8} watts m\textsuperscript{-2} (sub-A-class events). \par

\textit{NuSTAR} observations have allowed multiple detailed studies of sub-A-class events in active regions, as well as one paper concerned with three even smaller (\textit{GOES} $\sim$A0.01) quiet Sun brightenings \citep{Glesener_2017, Wright_2017, Cooper_2020, Kuhar_2018}. The spectra of events in \cite{Glesener_2017}, \cite{Kuhar_2018}, and \cite{Cooper_2020} were best fit by isothermal spectral models throughout their evolution, though the \cite{Glesener_2017} microflare displayed some high-energy excess over this fit during the impulsive phase. Pre-flare, post-flare, and decay phase spectra of the event presented in \cite{Wright_2017} were also best fit by a single thermal model, but the addition of a second higher-temperature thermal model was required to account for high energy excess during its impulsive phase. \par 

Additionally, two papers discuss slightly larger events. One considers an A1 class microflare that is the first observed by both \textit{NuSTAR} and the Interface Region Imaging Spectrograph (\textit{IRIS}), making it the first event at this scale where HXR coronal emission has been compared with corresponding cooler chromospheric UV emission as seen by \textit{IRIS} \citep{Hannah_2019}. The other presents a \textit{GOES} A7.7 \footnote{In \cite{Glesener_2020} the background-subtracted \textit{GOES} class for this event was reported as A5.7; the difference is due to the use of re-processed \textit{GOES} data released in May 2020 for this study.} event that is the first \textit{NuSTAR} microflare to show clear evidence of nonthermal emission \citep{Glesener_2020}. This last event occurred alongside several other microflares during a 2017 August 21 \textit{NuSTAR} observation, and is also discussed in this paper. \par

Limited \textit{NuSTAR} solar observing time means that full statistical A-class or sub-A-class microflare studies will have to wait for the introduction of a solar-dedicated focusing HXR mission. However, it is still valuable to strive for a more systematic understanding of these uniquely faint solar brightenings than we can gain from single-event studies. This motivates the analysis of the eleven events presented here.  \par


\subsection{Overview of Events}

On 2017 August 21, \textit{NuSTAR} observed a solar active region for an orbit of around an hour (\textit{NuSTAR} observation IDs 20312001001 and 20312002001). This observation was granted in conjunction with the ``Great American Eclipse" and ended with the eclipse of \textit{NuSTAR}'s FOV on the Sun by the Moon. Before the eclipse, four microflares of (background subtracted) \textit{GOES} A-class or below occurred in the \textit{NuSTAR} FOV, all originating within the targeted active region (NOAA designation AR 12671). The evolution of emission during the single orbit is shown in the top left panel of Figure \ref{fig:orbits}. \par

On 2018 May 29, \textit{NuSTAR} observed two solar active regions over the course of five orbits, each around an hour in duration. During this time, \textit{NuSTAR} recorded HXR emission from seven microflares, also all A-class or below. Six of these events were initially identified by visual inspection of \textit{NuSTAR} lightcurves. The last (may1917) was identified after a more rigorous method for identifying transients was applied. This set a series of conditions on the derivative of the \textit{NuSTAR} lightcurve to identify flare-like local maxima, based on the flare-finding algorithm used in \citet{2008ApJ...677.1385C} for a statistical study of \textit{RHESSI} microflares. All of the May events occurred during the first three orbits (\textit{NuSTAR} observation IDs 80410201001, 80410201002 and 80410201003) and within the same active region (NOAA designation AR 12712). \textit{NuSTAR} lightcurves for these three orbits are shown in the remaining panels of Figure \ref{fig:orbits}. \par

\begin{figure*}
\centering
\includegraphics[width=\textwidth, scale=1]{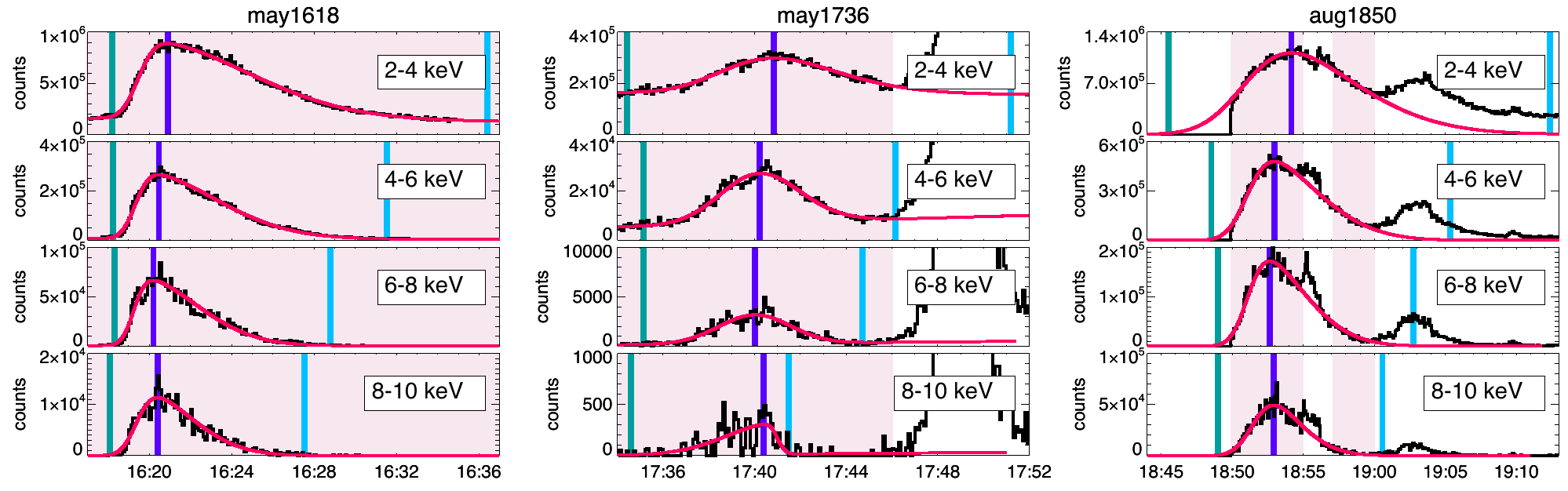}
\vspace{-0cm}
\caption{\textit{NuSTAR} lightcurves in four energy ranges during three of the eleven events (livetime-corrected, 6s binning), showing one `ideal' case, and two events where properties of the data or event caused challenges. Fitting intervals are shaded pink, fitted models (red) are plotted over \textit{NuSTAR} data, and the extracted start, peak, and end times (teal, purple, blue) are marked.  The left panels show a microflare (may1618) with a smooth, impulsive profile in all energy ranges (the `ideal' case). Center panels show the handling of a smaller event (may1736) with comparatively lower statistics available from 8-10 keV, and where the fitting interval is cut off by the rise of the next microflare. The right panels show an event (aug1850) that begins before the start of the \textit{NuSTAR} data interval and has a bump-like feature after the peak that prompted further trimming of the fitting interval to achieve a reasonable result.}
\label{fig:three_profiles}
\vspace{-0cm}
\end{figure*}

\textit{NuSTAR} observation of this particular region was motivated by the opportunity for co-observation with the \textit{Hi-C 2.1} sounding rocket, the flight of which occurred between two of the \textit{NuSTAR} observation intervals. Other co-observing instruments included the \textit{IRIS} high-resolution UV slit spectrometer, the \textit{Hinode} X-Ray Telescope (XRT) and Extreme-Ultraviolet Imaging Spectrometer (EIS), as well as the Atmospheric Imaging Assembly (AIA) aboard NASA's \textit{Solar Dynamics Observatory (SDO)} \citep{AIA_instrument}. Detailed analysis of AR 12712 during the quiescent interval of the \textit{Hi-C 2.1} flight is presented in \cite{Warren_2020}. The opportunity to incorporate results from \textit{Hinode} and \textit{IRIS} together with the \textit{NuSTAR} dataset is noted as an exciting area of future investigation. 
\par


\section{Temporal Evolution}\label{time}

The temporal structure of HXR emission in large flares (\textgreater B-class) is commonly impulsive, exhibiting a fast rise followed by a gradual fall. This is understood to imply an initial rapid release of energy to plasma heating and/or particle acceleration, followed by a lengthier decay interval as the heated plasma cools back down to temperatures below those that emit in the HXR band \citep[e.g.][]{benzFO}. \par

The time profile of higher energy HXR emission is generally observed to be more impulsive than that of the lower energy emission (lower energy HXRs, or SXR emission and lower energies), and also to peak earlier in time. This is consistent with a transfer of energy from accelerated particle populations and smaller, hotter plasma volumes into heating of the surrounding chromospheric plasma, as well as with gradual cooling over time. Both impulsivity and differential peak times between energy ranges are considered part of the ``standard" flare model \citep[e.g.][]{benzFO}), and consistent observation of them in microflare events would support the idea that the evolution of events at this scale is controlled by processes similar to those that lead to large flares.  \par

\begin{figure*}
\centering
\includegraphics[width=\textwidth, scale=1]{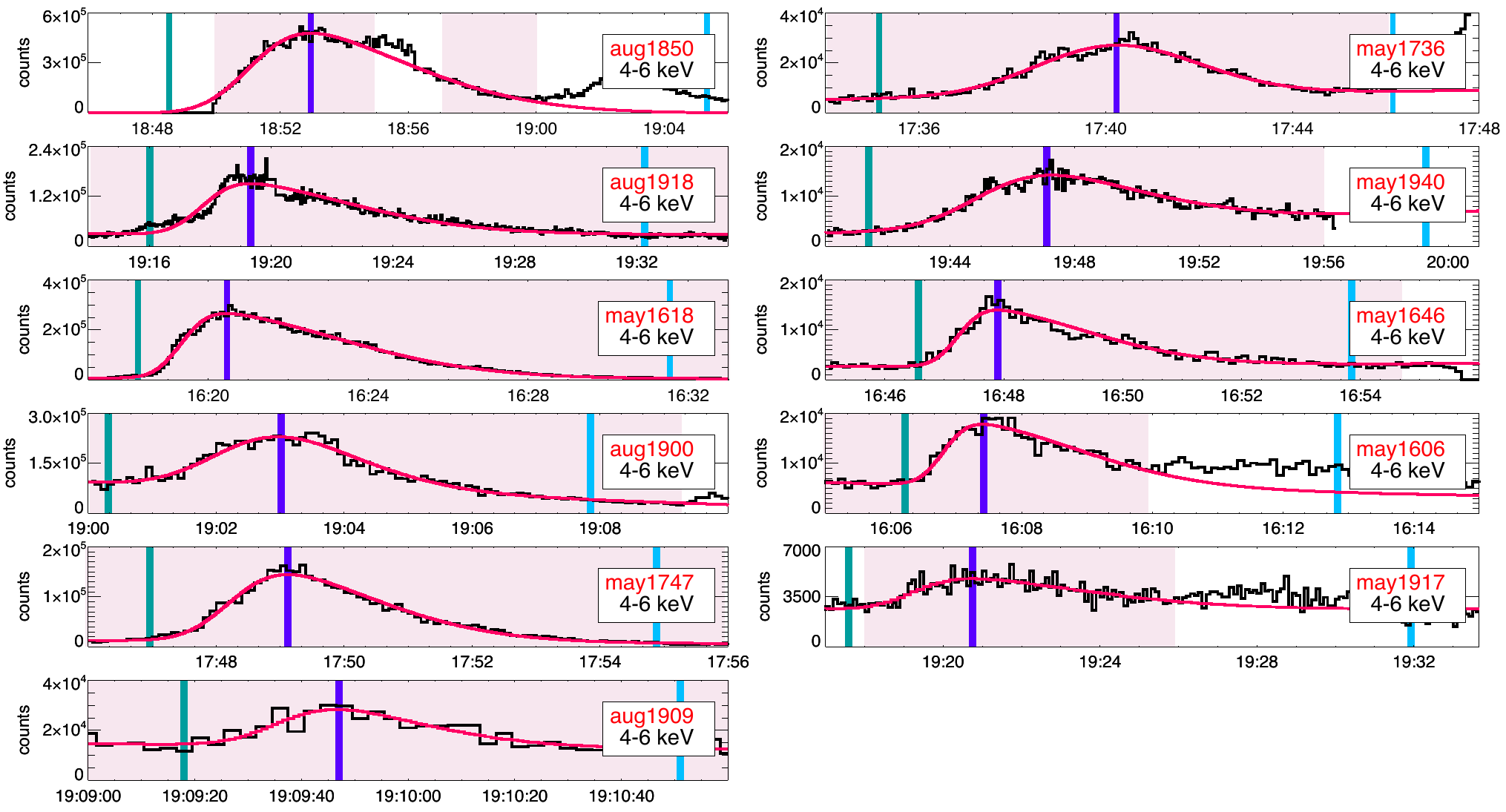}
\vspace{-0cm}
\caption{\textit{NuSTAR} 4-6 keV lightcurves over all flaring intervals (livetime-corrected, 6s binning), giving one fit example per event. Fitting intervals are shaded pink, fitted models (red) are plotted over \textit{NuSTAR} data, and extracted start, peak, and end times (teal, purple, blue) are marked. In some cases (aug1850, may1606, may1917), the fit interval was trimmed to minimize the effect of secondary bumps during the microflare decay, which are not well-fit by the single skewed gaussian. In others (aug1850 (start), may1736, may1940), the boundary of available \textit{NuSTAR} data or the rise of a new flare shorten the fit interval.}
\label{fig:profiles}
\vspace{-0cm}
\end{figure*}


\subsection{\normalsize Time Profile Analysis Method}

To examine these properties in the microflares considered here, four HXR energy bands were chosen within the observed \textit{NuSTAR} energy range (2-4 keV, 4-6 keV, 6-8 keV, 8-10 keV). An event asymmetry index (A\textsubscript{ev}) calculated from the rise and decay times (t\textsubscript{rise}, t\textsubscript{decay}) was used to examine the impulsivity in all 44 cases (4 energy bands $\times$ 11 events). This index was previously utilized to characterize the events in a \textit{RHESSI} microflare study \citep{2008ApJ...677.1385C}, following the example of \cite{Temmer_2002}. It is given as,
\begin{equation}
  A_{ev} = \frac{t_{decay} - t_{rise}}{t_{decay}+t_{rise}} 
\end{equation}
with a resulting value greater than zero implying an impulsive event. \par

Time profiles were created, including all \textit{NuSTAR} emission (FPMA, FPMB summed) observed in each energy range, integrated over selected regions. For May 2018, the regions chosen encompassed the full active region (AR 12712), which involved a relatively compact set of loops. In contrast, the August 2017 active region (AR 12671) was more structurally complex and highly elongated. In order to isolate microflare-specific temporal behavior, the regions chosen for August microflares included only the western half of the active region, the location of all four flare sites (see Figure \ref{fig:img} for \textit{NuSTAR} microflare emission plotted over AIA active region context data). \par

The time profiles included livetime-corrected \textit{NuSTAR} counts from several minutes before and after any flare emission was noticeable by eye, binned in 1s intervals. For a few events, the microflare either began or ended outside of the period of \textit{NuSTAR} observation, in which case as much of the flare-time interval was included as possible. \par

An automated method was developed to extract flare start, peak, and end times from each time profile. A model composed of a linear combination of skewed gaussian and linear functions (to represent flaring and background emission, respectively) was fit to each time profile, using the LMFIT Python package \citep{newville_matthew_2014_11813}. The skewed gaussian model was chosen for its ability to flexibly fit both impulsive and non-impulsive time profiles. The combination of the two functions requires six parameters to be fit (for the skewed gaussian: the center, width ($\sigma$), amplitude, and skewness ($\gamma$); for the linear component: the slope and intercept). \par

Fit quality was observed to be sensitive to the choice of initial conditions, so the fitting process was repeated iteratively for an array of initial conditions for three of the fit parameters (the gaussian center, $\sigma$, and amplitude). Optimal sets of initial conditions were found (those resulting in the best fit, with goodness-of-fit determined via the chi-squared value). Using these, best fit parameters were extracted. This was repeated for each of the 44 time profiles. Figure \ref{fig:three_profiles} shows the available \textit{NuSTAR} data, the interval used for fitting, and the fit results in all four energy ranges for three selected events, while Figure \ref{fig:profiles} shows the 4-6 keV fit and data for every event considered. \par

The peak time was defined as the time of the maximum of the resulting model function. Start and end times were defined as times when the integral of the skewed gaussian model component (with the background component removed) was equal to 0.1\% and 99.9\% (respectively) of its value when evaluated over the full input duration. These thresholds are arbitrary, but produced reasonable start/stop times in comparison to what was apparent to the eye for each event where we had a clean observation (\textit{NuSTAR} data over the full duration, with no overlapping events - see Figure \ref{fig:profiles}, microflares aug1918, may1618, aug1900, may1747, aug1909, may1646). The resulting degree of impulsivity was not strongly dependent on the exact values of the thresholds. \par

For 4/11 microflares (aug1850, may1646, may1736, may1940), the full evolution of the event was not captured in the \textit{NuSTAR} data: the interval was cut short by either the boundaries of the observation, or by another flaring event occurring shortly after. With the use of fit results, start/end times were estimated even beyond the available \textit{NuSTAR} data in these cases. Additionally, three events (may1850, may1606, may1917) contained bumps in the decay interval which distorted fit results, pushing end times well beyond a value that seemed physical. For these events, the interval used for fitting was manually trimmed to avoid including these features.  \par

Uncertainty in peak times was dependent on LMFIT output 1-$\sigma$ standard errors in both the center and $\gamma$ of the distribution, while uncertainties in start and stop times were additionally dependent on the error in  $\sigma$. To extract uncertainties, fits were iteratively re-run for each time profile wherein the center, $\gamma$, and $\sigma$ were randomly assigned to values within their output error range each time, and then held fixed while the other three parameters remained free. Peak, start, and end time uncertainties were taken as the standard deviation of their resulting values. \par 


\subsection{ Peak Times \& Impulsivity }\label{resz}

\begin{deluxetable}{|c|c|c|c|c|}
\tabletypesize{\footnotesize}
\tablecaption{Event Asymmetries (A\textsubscript{ev}), Shaded to Indicate Sign of Values \label{times2}} 
\tablehead{ \colhead{\textbf{Event}} &  \colhead{\textbf{2-4 keV}} &   \colhead{\textbf{4-6 keV}} &   \colhead{\textbf{6-8 keV}} &  \colhead{\textbf{8-10 keV}}}
\startdata
\hline
aug1850 & \cellcolor{green!25}0.36$\pm$0.08 & \cellcolor{green!25}0.47$\pm$0.13 & \cellcolor{green!25}0.47$\pm$0.14 & \cellcolor{green!25}0.32$\pm$0.23 \\
\hline
aug1918 & \cellcolor{green!25}0.55$\pm$0.02 &\cellcolor{green!25} 0.59$\pm$0.02 &\cellcolor{green!25} 0.78$\pm$0.02  & \cellcolor{green!25}0.79$\pm$0.10  \\
\hline
may1618 & \cellcolor{green!25}0.70$\pm$0.01 & \cellcolor{green!25}0.67$\pm$0.01  & \cellcolor{green!25}0.64$\pm$0.04 &\cellcolor{green!25} 0.51$\pm$0.10 \\ 
\hline      
aug1900 & \cellcolor{green!25}0.43$\pm$0.02  &\cellcolor{green!25} 0.28$\pm$0.03 & \cellcolor{green!25}0.35$\pm$0.04  & \cellcolor{green!25}0.66$\pm$0.09  \\
\hline      
may1747 &  \cellcolor{green!25}0.48$\pm$0.02  &\cellcolor{green!25} 0.46$\pm$0.02 &\cellcolor{green!25} 0.37$\pm$0.10 & \cellcolor{green!25}0.31$\pm$0.22 \\
\hline   
aug1909 & \cellcolor{green!25}0.69$\pm$0.10  & \cellcolor{green!10}0.38$\pm$0.75 & \cellcolor{green!10}0.90$\pm$4.6 & \cellcolor{green!10}0.54$\pm$0.63 \\
 \hline
may1736 & \cellcolor{green!25} 0.23$\pm$0.03  & \cellcolor{green!10}0.07$\pm$0.07  & \cellcolor{green!10}-0.02$\pm$0.17  &-0.68$\pm$0.06 \\
\hline       
may1940 & \cellcolor{green!25}0.46$\pm$0.03  & \cellcolor{green!25}0.36$\pm$0.03 & \cellcolor{green!25}0.32$\pm$0.06  & \cellcolor{green!10}0.22$\pm$0.36  \\
\hline  
 may1646 & \cellcolor{green!25}0.69$\pm$0.04  &\cellcolor{green!25} 0.63$\pm$0.04 & \cellcolor{green!25}0.46$\pm$0.14  &\cellcolor{green!25} 0.70$\pm$0.29 \\
\hline        
may1606 & \cellcolor{green!25}0.62$\pm$0.05  & \cellcolor{green!25}0.64$\pm$0.03 & \cellcolor{green!25}0.80$\pm$0.09  & \cellcolor{green!25}0.95$\pm$0.85 \\ 
\hline
may1917 &  \cellcolor{green!25}0.86$\pm$0.01 &  \cellcolor{green!25}0.56$\pm$ 0.07& \cellcolor{green!25} 0.28$\pm$0.28 &  \cellcolor{green!10} X \\
\hline
\hline
\hline
& Color &\cellcolor{green!25} Impulsive &Consistent \cellcolor{green!10} & Non-Impulsive\\
& Key: &\cellcolor{green!25} (A\textsubscript{ev}\textgreater\ 0) &\cellcolor{green!10} With Either  & (A\textsubscript{ev} $\leq$ 0) \\ 
\enddata
\tablecomments{A\textsubscript{ev} was calculated using t\textsubscript{rise} and t\textsubscript{decay} found independently in each energy range. Due to poor statistics, the 8-10 keV range for the faintest event (may1917) was excluded from analysis.}
\end{deluxetable}

\begin{figure}
\includegraphics[width=\linewidth]{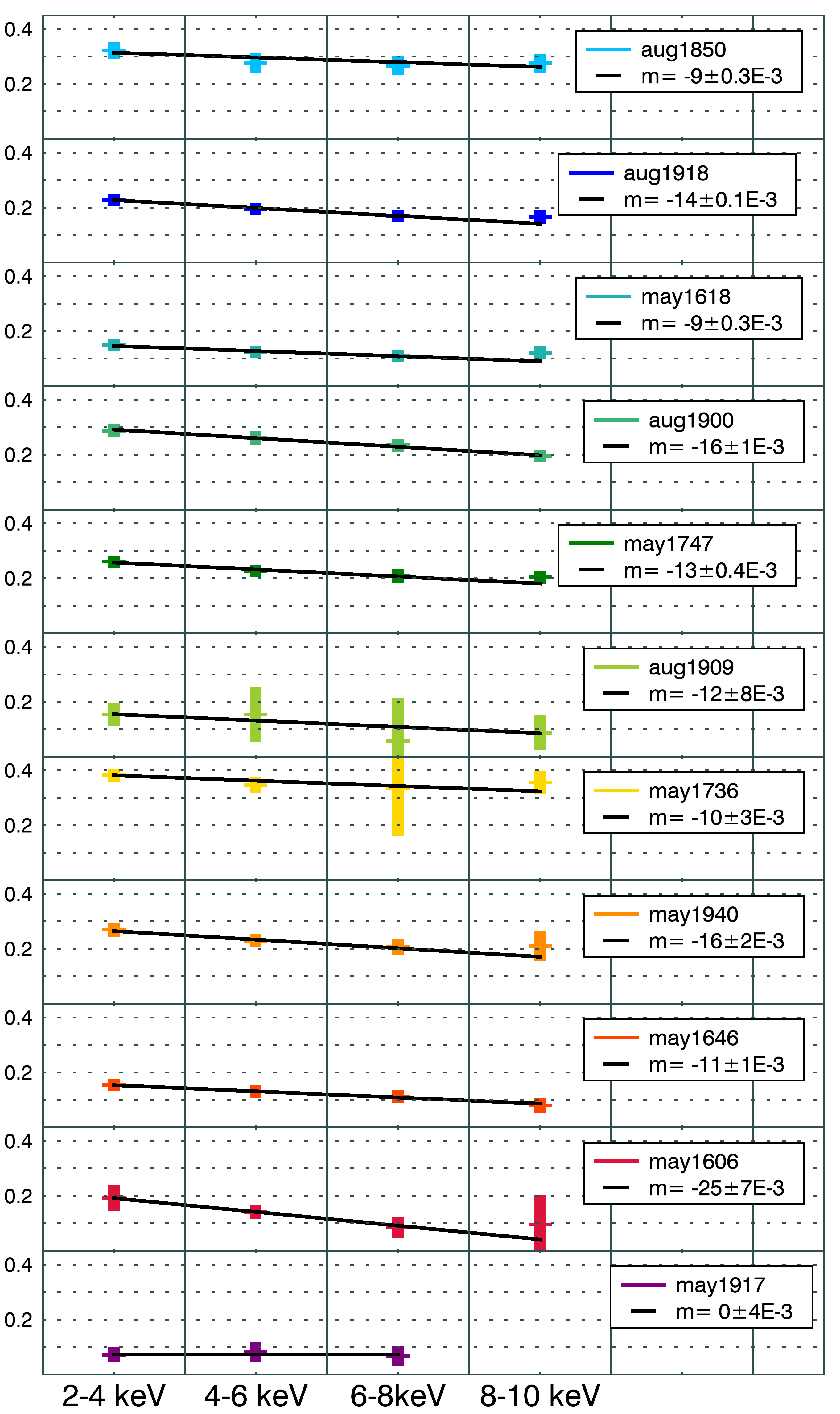}
 \caption{Microflare peak times are shown, normalized over the 2-4 keV event duration such that a value of 0 would imply the microflare peaks the moment it begins, while a value of 0.5 would imply a peak halfway through the duration. Peak times are shown in all four energy ranges, with error bars showing uncertainties for each. A linear fit is included for each event, and resulting slopes (m) are reported with 1-$\sigma$ uncertainties. The data are consistent with a negative slope only in 10/11 cases (a slope of zero was found for the faintest event, may1917, which had sufficient counts to be well-fit in only three energy ranges). This shows a trend toward earlier peak times in the higher energy ranges. Events are arranged from brightest (top) to faintest (bottom) by the maximum \textit{NuSTAR} count rate (livetime corrected and background subtracted) during each interval.}
\label{fig:layercake}
\end{figure}

Table \ref{times2} shows the event asymmetry index, A\textsubscript{ev}, for each microflare in each of the four energy ranges. Events are arranged from brightest (top) to faintest (bottom) considering the maximum \textit{NuSTAR} count rate during each (livetime corrected and background subtracted). The majority (36/43, dark green) of the time profiles are confirmed to be impulsive, and 6 more (light green) are consistent with either an impulsive or non-impulsive evolution. \par

One event (aug1909) was consistent with both impulsive and non-impulsive profiles in 3/4 energy ranges. This was the shortest-duration microflare (\textless2 minutes). While start, peak, and end time uncertainties in this event were not notably larger than those found for others, they are larger in proportion to the flare duration, leading to large uncertainty ranges in A\textsubscript{ev} in the higher energies. The lack of confirmed impulsivity in these time profiles therefore reflects a limitation of the available statistics. \par

\begin{figure}
\includegraphics[width=\linewidth]{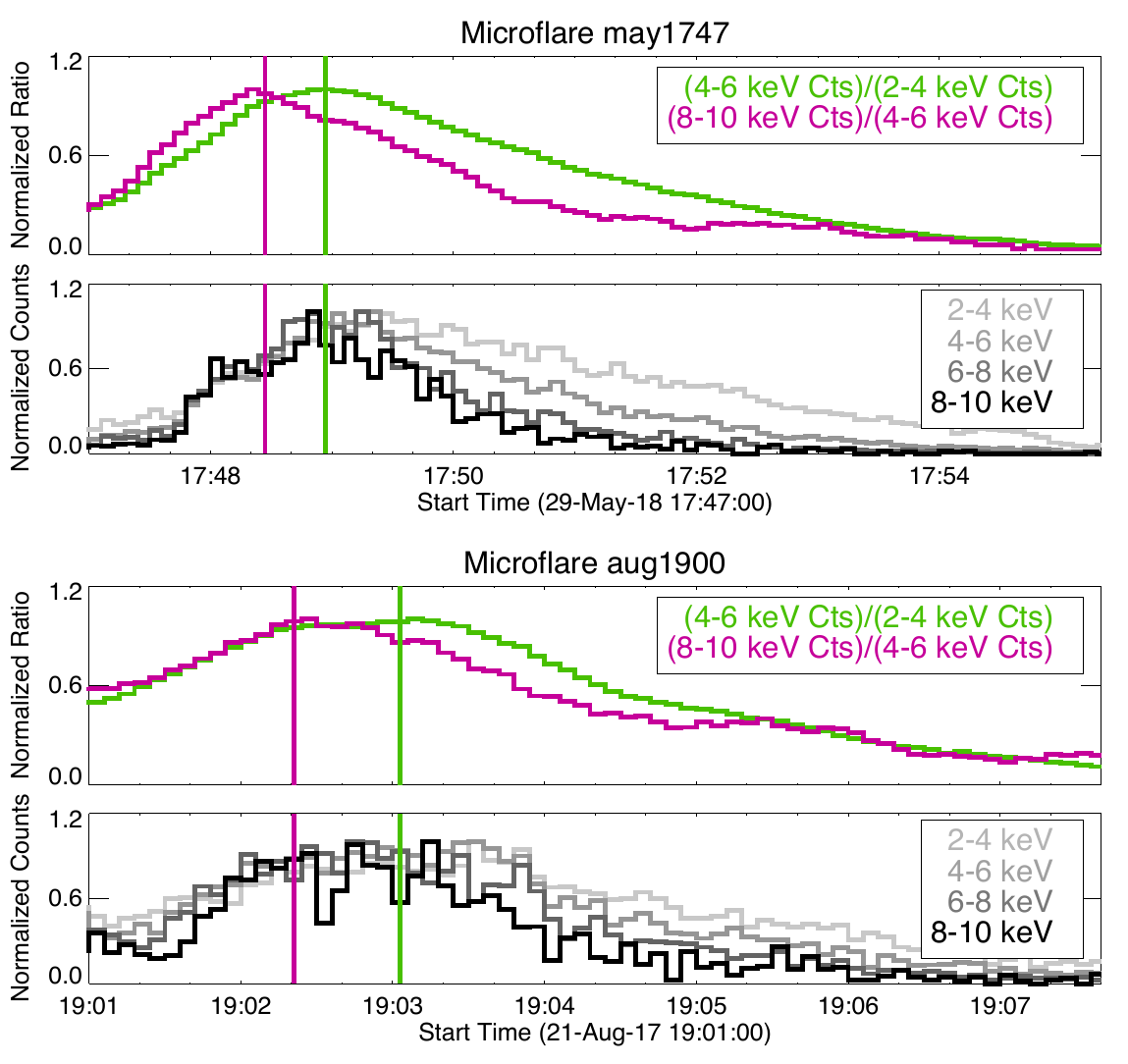}
 \caption{ Time evolution of ratios between two sets of \textit{NuSTAR} energy ranges are shown for two example events, a proxy for temperature. Ratio curves have been smoothed over a 10s interval, and their peak times are marked with color coded vertical lines and extended down for comparison with normalized \textit{NuSTAR} emission in four energy ranges. In the top event (may1747), the two ratios peak before the peaks in emission in any of the \textit{NuSTAR} channels, while they straddle the \textit{NuSTAR} peaks in the bottom event (aug1900). }
\label{fig:tper}
\end{figure}

The reported values of A\textsubscript{ev} represent the location of the time profile peak within the event duration in each energy range. It is also interesting to compare time profile peaks in each energy range to the full event duration (defined as the interval from the earliest start to the latest end found in any of the energy ranges for each event; in all 11 microflares this was equal to the 2-4 keV event duration). The peak times of emission in all energy ranges for each microflare are shown in Figure \ref{fig:layercake}. In order to visually compare between energy ranges for all microflares simultaneously, the peak times were normalized over each 2-4 keV event duration. Linear trends in peak time across the four energy ranges were calculated for each microflare. 10/11 resulting trendlines had negative slopes, confirming observation of the large-flare property of earlier peak times in higher energy emission.  \par


\subsection{Hardness Ratios}\label{hardrad}

The differential flux spectrum of thermal bremsstrahlung from a volume of plasma is dependent on the electron and ion densities (n\textsubscript{e}, n\textsubscript{i}), as well as the temperature (T) of the plasma. It is given as a function of emitted energy, $\epsilon$, as

\begin{equation}
F(\epsilon) \approx 8.1\times10^{-39} \int_{V} \frac{\exp(\frac{-\epsilon}{k_{B}T})}{T^{\frac{1}{2}}} n_{i}n_{e} \dd{V} 
\end{equation}

\noindent
( keV s\textsuperscript{-1} cm\textsuperscript{-2} keV\textsuperscript{-1} ) where factors on the order of 1 have been neglected, and the integral is taken over the volume of emitting plasma \citep{2005psci.book.....A}. The ratio of this flux at two different energies can be shown to be a monotonically increasing function of T. \par

With sufficient knowledge of instrument response, this relationship can be used to determine the evolution of flare temperature in absolute terms (as was done for a large population of \textit{GOES} flares in \cite{2012ApJS..202...11R}). \textit{NuSTAR}'s energy resolution allows for flare temperatures to be more accurately extracted from spectroscopy (see Section \ref{spec}). However, the need to include enough counts to make spectral fitting meaningful limits the temporal resolution possible when examining the evolution of flare plasma parameters over the course of an event. Hardness ratios (ratios between counts in higher and lower \textit{NuSTAR} energy ranges) do not have this limitation. \par

\begin{deluxetable}{|c|c|c|}
\tabletypesize{\small}
\tablecaption{Hardness Ratio Peaktimes (Fraction of Flare Duration) \label{erp}} 
\tablehead{ \colhead{Event} & \colhead{R\textsubscript{4/2}}   & \colhead{R\textsubscript{8/4}}  }
\startdata 
aug1850*  &    0.264$\pm$0.006 * &     0.277$\pm$0.008 * \\
\hline
aug1918 &      0.183$\pm$0.015 &      0.178$\pm$0.020\\
\hline
may1618 &    0.096$\pm$0.025 &     0.111$\pm$0.036\\
\hline
aug1900 &      0.271$\pm$    0.026 &      0.198$\pm$    0.023\\
\hline
may1747 &    0.203$\pm$0.019 &    0.165$\pm$0.022\\
\hline
aug1909 &   0.038$\pm$0.139 &     0.125$\pm$0.097\\
\hline
may1736 &   0.339$\pm$0.019 &     0.296$\pm$0.044\\
\hline
may1940 &      0.226$\pm$0.012 &      0.184$\pm$0.039\\
\hline
may1646  &  0.148$\pm$0.036 &     0.102$\pm$0.041\\
\hline
may1606 &      0.173$\pm$0.028 &      0.161$\pm$0.038\\
\hline
may1917 &    0.081$\pm$0.016 &  0.043$\pm$0.071\\
\enddata
\tablecomments{ The significant nonthermal contribution to emission in aug1850 (see Section \ref{spec}) complicates interpretation of the ratio peak as a temperature peak in this case.}
\end{deluxetable}

Two different hardness ratios were examined in these events: R\textsubscript{4/2} (ratio of 4-6 keV emission and 2-4 keV emission), and R\textsubscript{8/4} (ratio of 8-10 keV emission and 4-6 keV emission). Figure \ref{fig:tper} shows both ratios as a function of time during two example events, with normalized \textit{NuSTAR} emission in all four energy ranges included for context. The hardness ratios are normalized over the flaring interval for visual convenience in comparing between R\textsubscript{4/2} and R\textsubscript{8/4}. \par

These events are representative of the population of microflares, all of which showed ratios with structure similar to that of the regular \textit{NuSTAR} time profiles, peaking either simultaneously or earlier in time. The exceptions to this were smaller events, where limited statistics in the 8-10 keV energy range challenged the interpretation of R\textsubscript{8/4}. 

Table \ref{erp} gives peak times in each of the hardness ratios for each microflare, events again arranged by magnitude of peak \textit{NuSTAR} counts (livetime corrected, background subtracted). The ratio peak times are reported as fractions of the full 2-4 keV event duration in each case. Uncertainties were found by applying a range of different smoothing intervals to each ratio curve before taking the maximum, and using the standard deviation of the resulting peak times as the reported uncertainty. The uncertainty associated with the choice of smoothing interval was seen to dwarf that due to the inherent statistical uncertainty of the \textit{NuSTAR} data. For both ratios in all eleven microflares, the peak occurs in the first half of the event. The mean values of the peaks (times of maximum microflare temperature assuming an isothermal emitting plasma) are 0.176$\pm$0.034 (R\textsubscript{4/2}) and 0.156$\pm$ 0.043 (R\textsubscript{8/4}) when averaged over all events except aug1850.\par


\subsection{Neupert Effect }

The Neupert effect describes the tendency for flaring HXR or microwave emission to show correlation with the derivative of the lightcurve of emission in lower energy ranges, as noted in \cite{1968ApJ...153L..59N}. Observation of this property is interpreted to support the idea that plasma heating resulting in EUV and SXR emission is caused by the deposition of energy by beams of nonthermal accelerated electrons, which are in turn the source of emission in the HXR band \citep{Dennis_1993}. \par

\begin{figure}
\includegraphics[width=\linewidth]{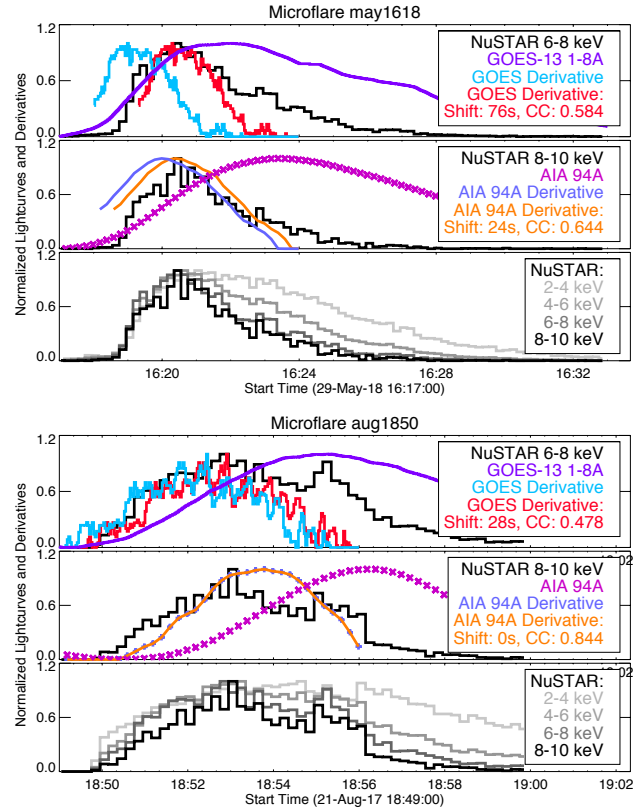}
\caption{Correlation results between \textit{NuSTAR} and the derivatives of lower-energy emission for two events. For each event, three plots are shown (Top: \textit{GOES} derivatives over the cross-correlation interval, with full flare duration \textit{NuSTAR} and raw \textit{GOES} lightcurves for context. Middle: the same is shown comparing \textit{NuSTAR} and AIA. Bottom: normalized \textit{NuSTAR} emission is shown in all four energy ranges over the flare duration. Legends on the \textit{GOES} and AIA lightcurves include the correlation coefficient (CC) and associated shift.). \textbf{Upper Event:} (may1618) time profiles are best correlated with AIA 94 \AA\ and \textit{GOES} 1-8 \AA\ derivatives when they are shifted forward in time (not consistent with the Neupert effect). \textbf{Lower Event:} (aug1850, an event confirmed to contain nonthermal emission) The 8-10 keV time profile is best correlated with the AIA 94 \AA\ derivative with no shift, and the rise in 6-8 keV is well correlated with the \textit{GOES} 1-8 \AA\ derivative even before a shift is applied to maximize the mathematical correlation (consistent with the Neupert effect).}
\label{fig:n_e}
\end{figure}

In order to look for evidence of this, co-temporal lower energy emission was examined in conjunction with higher-energy (6-8 keV, 8-10 keV) \textit{NuSTAR} time profiles. SXR emission was taken from the \textit{GOES} 1-8 \AA\ passband, and EUV from the \textit{SDO}/AIA 94 \AA\ channel, the latter spatially integrated over the relevant active regions. This meant that, in total, 44 pairs of lightcurves were examined (2 \textit{NuSTAR} energy ranges $\times$ 11 microflares $\times$ 2 lower-energy instruments). \par

AIA and \textit{GOES} time profiles were smoothed over 2 minute boxcar intervals before their derivatives were taken, with the aim of highlighting longer-term temporal structure over background fluctuations. This was modified in the case of aug1909, where the \textit{GOES} and AIA emission were smoothed over 1 minute instead due to the \textless\ 2 minute event duration. Intervals were selected over which to compare \textit{NuSTAR} emission with the \textit{GOES} or AIA derivatives, including only the times where the \textit{GOES} or AIA derivatives were non-negative. \textit{NuSTAR} data were binned to match the cadence of the lower-energy instruments, and emission in each of the instruments was normalized. \par

Cross correlation between each pair of lightcurves was computed using the \texttt{C\_CORRELATE} function in IDL, reporting the maximum correlation coefficient found and its associated lag. Figure \ref{fig:n_e} shows the best-correlated result (from either the 6-8 or 8-10 keV comparison) for both \textit{GOES} and AIA for two events. \par

In 18/22 AIA comparisons and 17/22 \textit{GOES} comparisons, the best correlation between \textit{NuSTAR} emission and the lower energy derivative was found after a positive shift of the derivative in time (positive lag), implying the lower-energy derivative peaks earlier in time than the \textit{NuSTAR} emission. An example of this can be seen in the upper panel of Figure \ref{fig:n_e}. \par 

If an increase in the amount of plasma emitting in a given SXR or EUV energy range were the result of heating by a nonthermal particle population, the derivative would be expected to peak at the same time as (or later than) the nonthermal emission. In contrast, events best-correlated with a positive lag are explainable by entirely thermal emission in both the HXR and SXR/EUV ranges, and are not consistent with the Neupert effect. This appears to be the dominant behavior among this population. \par

Considering the events not best-correlated when a positive lag is applied, the majority were cases in which the correlation between \textit{NuSTAR} and the lower energy derivative was weak, or seemed unphysical. Some of the smallest events were faint enough to be difficult to discern in the \textit{GOES} light curves, but even some that were visible did not display a strong correlation. This is likely indicative of a more complex physical situation than is assumed by either the Neupert effect or the simple thermal scenario described above. \par

Microflare aug1850, which was confirmed to involve significant nonthermal emission in \cite{Glesener_2020}, is the only event to show behavior consistent with the Neupert effect with the use of this method. Specifically, the AIA derivative is best-correlated with both the \textit{NuSTAR} 6-8 keV and 8-10 keV time profiles when no lag is applied at all (see lower panel of Figure \ref{fig:n_e}). The \textit{GOES} derivative, while requiring a small positive shift to achieve the best mathematical correlation, does also qualitatively appear well-correlated with the rise of the \textit{NuSTAR} emission in both energy ranges without being shifted at all (6-8 keV correlation coefficients: 0.399 (no shift),  0.478 (28s shift); 8-10 keV correlation coefficients: 0.357 (no shift), 0.457 (58s shift)). \par


\section{Spatial Properties}\label{space}

\begin{figure*}
\includegraphics[width=\textwidth, scale=1]{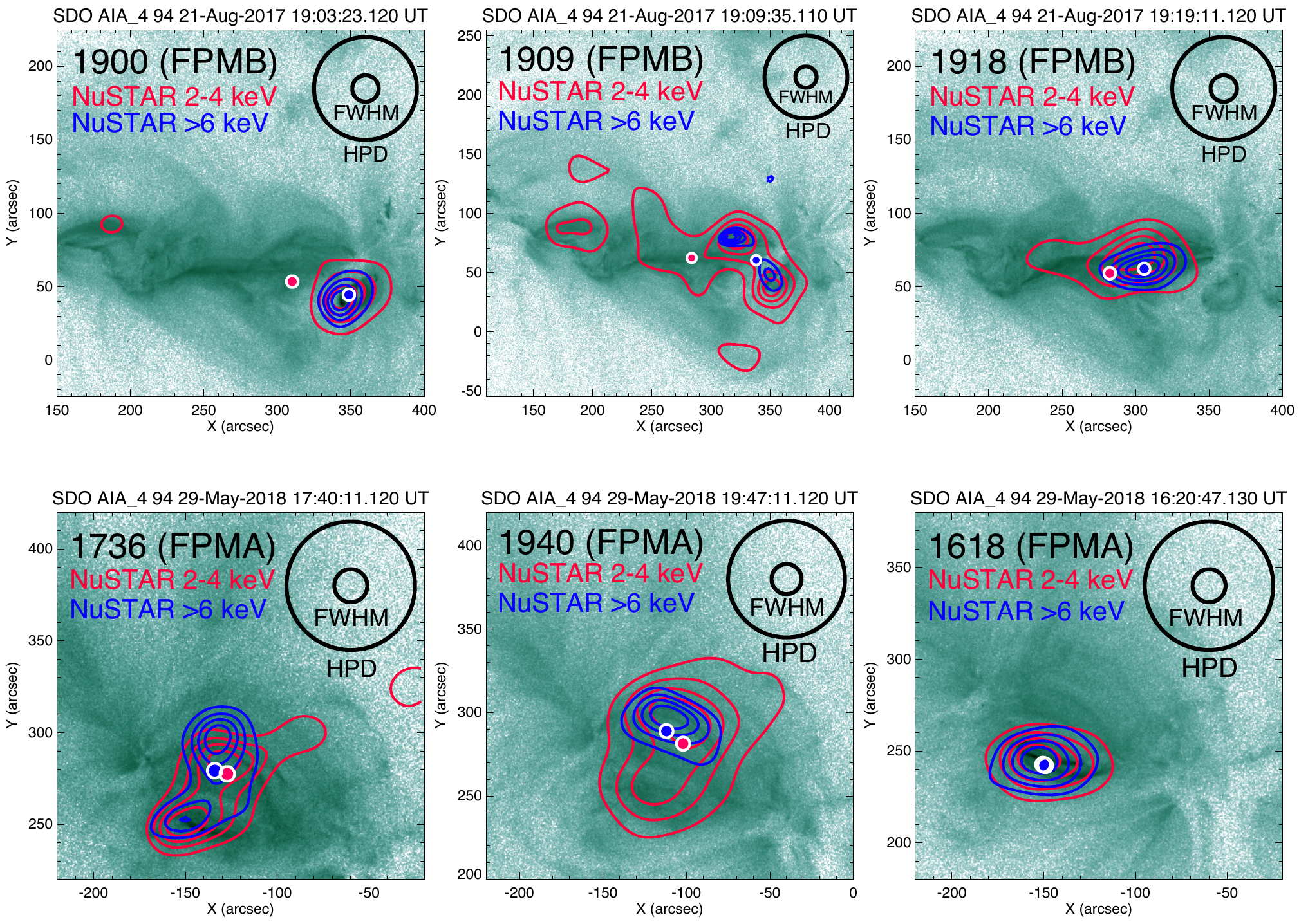}
 \caption{\textit{NuSTAR} contours (20, 40, 60, 80\% of counts) are shown over AIA 94\AA\ context images during six of the eleven microflares. \textit{NuSTAR} 2-4 keV and \textgreater 6 keV emission during each event has been integrated over stable pointing intervals and deconvolved over flare- and energy range- specific numbers (i) of iterations. Emission centroids in each energy range are also marked. Circles with diameters equal to the \textit{NuSTAR} HPD ($\lesssim$70$"$ in this energy range) and FWHM (18$"$ ) are shown for visual reference \citep{2015ApJS..220....8M}. \textbf{Upper Panels:} Three of the August 2017 microflares are shown, each with unique morphology involving different parts of the complex, multi-loop structure of AR 12671 (see \cite{Glesener_2020} for images of microflare aug1850). No pre-flare background has been removed, as no times during the \textit{NuSTAR} observation of this region could be considered quiescent. The same number of iterations were used for deconvolution of each (2-4 keV: i=200, \textgreater 6 keV: i=100). \textbf{Lower Panels:} In contrast, emission in all May 2018 events was dominated by the contributions of two similarly shaped structures (an `upper' and `lower' set of loops). The lower left panel shows microflare may1736 (no pre-flare background has been removed in this case, 2-4 keV: i=100, \textgreater 6 keV: i=50). This event involved significant \textit{NuSTAR} emission from both the  `upper' and `lower' loop structures. The center panel shows may1940, an event primarially involving the `upper' loop structure (2-4 keV: i=100, \textgreater 6 keV: i=50), while the right (may1618)  shows one dominated by the `lower' (2-4 keV: i=200, \textgreater 6 keV: i=100). May1618 was the brightest of the May 2018 microflares, and has had background emission removed. No significant energy-dependent difference in \textit{NuSTAR} centroid was found for this event.}
\label{fig:img}
\end{figure*}
\textit{NuSTAR}'s imaging capabilities allow for comparison between the spatial distribution of observed HXR emission with that of EUV emission observed by \textit{SDO}/AIA. \textit{NuSTAR}'s 18$"$  angular resolution (FWHM) means that structure on the scale of larger active region loops can be resolved, though much of the finer loop spatial structure visible in AIA is not. As a first step in investigating spatial properties of the observed emission, \textit{NuSTAR}'s pointing stability was examined over each flaring interval. In 10/11 microflares, the CHU combination and pointing were stable over the entire flare (may1606, may1736, may1917, may1940, aug1900, aug1909), or over the rise and peak times (may1646, may1618, may1747, aug1918). For these events, data from the dominant (or rise/peak) CHU combination were used to make images, reducing event duration in some cases (the same intervals were later used for spectroscopy in Section \ref{spec}). For aug1850, multiple CHU changes occurred during the rise/peak, so the CHU with the largest effective exposure was chosen for imaging. \par 

Depending on the location of the detector chip gap in relation to an observed source, one FPM may be more ideal for imaging in any given observation. FPMA was better oriented during the May 2018 observation, while FPMB was better during all August 2017 events. To make images, \textit{NuSTAR} emission from one FPM was integrated in time over the CHU-stable intervals for each microflare. \textit{NuSTAR}'s point-spread function was then deconvolved over an event- and energy-range-specific number of iterations using the IDL procedure \texttt{max\_likelihood.pro}. \par

Because \textit{NuSTAR} is sensitive at temperatures similar to those that most strongly produce the Fe XVIII line (peak formation temperature of log(T)$\approx$6.9 \citep{2015A&A...582A..56D}), AIA Fe XVIII images can be used to approximate the most likely true center of \textit{NuSTAR} emission, reducing the instrument's inherent pointing uncertainty during solar observation. AIA Fe XVIII images were produced using an established linear combination of three channels (94\AA, 171\AA, and 211\AA) to isolate Fe XVIII emission \citep{2013A&A...558A..73D}. Differenced Fe XVIII images were then created (peak time in \textit{NuSTAR} 2-4 keV minus a pre-flare time). \par

Finally, for each flare, the deconvolved \textit{NuSTAR} contours were manually coaligned to the differenced Fe XVIII images. Figure \ref{fig:img} shows \textit{NuSTAR} emission from six microflares as contours over AIA 94\AA\ context. The number of deconvolution iterations used for each energy range in each event are given in the image caption. \par

\subsection{Spatial Complexity}

Differences in the centroid of flare-time emission in different HXR energy ranges could provide evidence of a plasma temperature gradient across the flare site. Alternatively, such differences could highlight spatially distinct thermal and nonthermal sources, such as the common scenario of nonthermal loop footpoint sources in large flares observed in conjunction with thermal emission from flare loops \citep[e.g.][]{benzFO}. To determine whether an event displays spatial complexity, background active region emission must first be subtracted from the flare-time images. This ensures the isolation of complexity within the microflare itself, rather than just characterization of a spatial difference between flare emission and (generally lower energy) emission from the surrounding active region. \par

Suitable quiet times for background subtraction were found for only two events (16:44-16:45 UT for may1618, shown in Figure \ref{fig:img}; 18:12-18:15 UT for may1747), as essentially no quiet times occurred during the August 2017 observation (see Figure \ref{fig:orbits}), and the remaining May 2018 microflares were faint enough that background subtraction resulted in poor statistics and significant non-physical distortion. After background subtraction, the \textit{NuSTAR} emission centroid was computed in the 2-4, 4-6, and \textgreater 6 keV energy ranges for each FPM, considering all pixels with values above 15\% of the maximum pixel value in each raw (not deconvolved) image. Differences in centroid between the two FPM in the full \textit{NuSTAR} energy range (all energies \textgreater 2 keV) were used as an estimate of uncertainty in the centroid measurements. Neither of the events displayed a difference between emission centroids in different \textit{NuSTAR} energy ranges larger than the estimated uncertainty. This is consistent with what is observed in AIA Fe XVIII, where both of these events showed dominant emission from just one feature.  \par


\section{\large Spectroscopy}\label{spec}

\begin{figure*}
\centering
\includegraphics[width=0.85\linewidth]{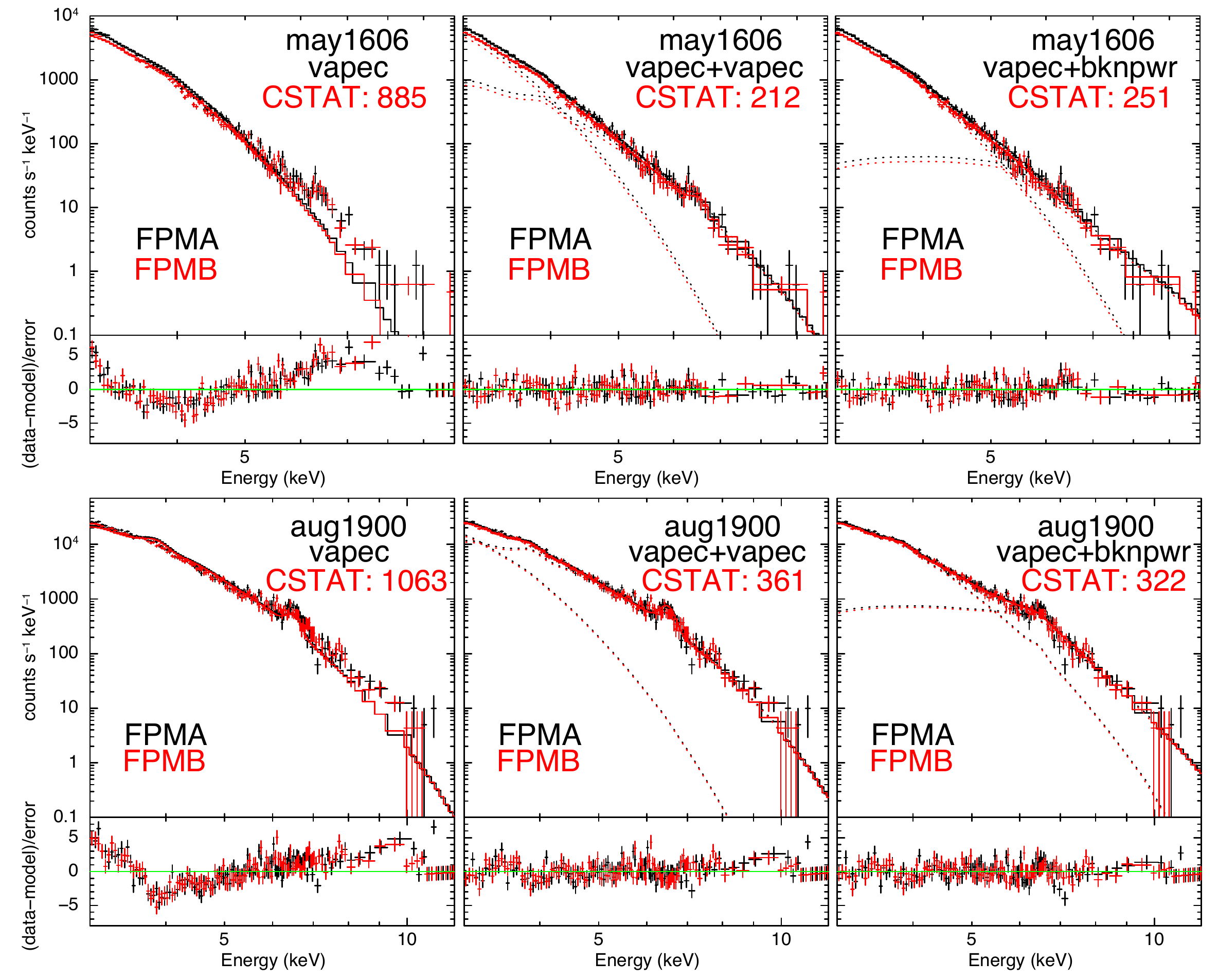}
\vspace{-0cm}
\caption{Spectra of may1606 (16:06–16:16 UTC, top row) and aug1900 (19:01–19:08 UTC, bottom row), using three different models (\texttt{vapec} (left), \texttt{vapec+vapec} (center), \texttt{vapec+bknpwr} (right)). For each example, livetime-corrected FPMA (black) and FPMB (red) count spectra are shown in each upper panel, while the lower panels show the error-normalized residuals. For may1606 (one of the faintest microflares), spectral fits were performed between 3-10 keV. The \texttt{vapec+vapec} model provides the strongest fit to the data. The significantly brighter aug1900 was fit over 3-12 keV, and required a gain correction (see Appendix \ref{gain}). The slope of the linear gain was freed while performing the \texttt{vapec+vapec} fit, and its resulting value (0.990) was applied as a fixed correction for the \texttt{vapec} and \texttt{vapec+bknpwr} fits. For this event, the \texttt{vapec+vapec} and \texttt{vapec+bknpwr}  fits were similar in quality. Thermal fit parameters for both events are reported in Table \ref{therm}, and \texttt{vapec+bknpwr} parameters are reported for aug1900 in Table \ref{nontherm}.}
\label{fig:spec_ex}
\vspace{-0cm}
\end{figure*}

\begin{deluxetable*}{cccccccc}
\tabletypesize{\footnotesize}
\tablecaption{Fit Thermal Parameters* (\texttt{vapec+vapec})  \label{therm}} 
\tablehead{ \colhead{\textbf{Event}} &  \colhead{\textbf{VAPEC1}} &   \colhead{\textbf{VAPEC2}} &   \colhead{\textbf{GAIN SHIFT}} & \colhead{\textbf{Loop Density}}&  \colhead{\textbf{Energy}}&  \colhead{\textbf{GOES}} & \colhead{\textbf{GOES}} \\
\colhead{\textbf{}} &  \colhead{T (MK)} &   \colhead{T (MK)} &   \colhead{} &  \colhead{(cm\textsuperscript{-3})}& \colhead{Thermal (erg)}&  \colhead{calc.} & \colhead{obs.} \\
\colhead{\textbf{}} &  \colhead{EM (cm\textsuperscript{-3})} &   \colhead{EM (cm\textsuperscript{-3})}  &   \colhead{} &  \colhead{}& \colhead{}&  \colhead{} & \colhead{}}
\startdata
aug1850* & $10.3\substack{+0.2\\ -0.2}$\   & & 0.972$\pm$0.007 (A) & $4.8\substack{+0.6 \\ -1}\times10^{9}$\ & $5.0\substack{+1.2\\ -0.7}\times10^{27}$\   & & A7.7 \\
& $5.6\substack{+0.3 \\ -0.3}\times10^{45}$\  & &  0.979$\pm$0.007 (B)  &   &  &\\
\hline
aug1918 & $4.1\substack{+0.1\\ -0.1}$\                                      &  $10.0\substack{+0.02\\ -0.1}$\                                                    &  0.984$\pm$0.003 & $5.2\substack{+1 \\ -1}\times10^{9}$\  & $9.1\substack{+3\\ -2}\times10^{28}$\  &  A$4.7\substack{+0.1\\ -0.7}$\ & A4.5 \\
 & $2.8\substack{+0.1 \\ -0.1}\times10^{47}$\  &  $4.4\substack{+0.2 \\ -0.1}\times10^{45}$\   &  & &   &  &\\
\hline
may1618 & $4.1\substack{+0.2\\ -0.1}$\                                            & $10.0\substack{+0.03\\ -0.03}$\  &  0.977$\pm$0.002                & $5.6\substack{+2 \\ -2}\times10^{9}$\  & $4.3\substack{+4\\ -1}\times10^{28}$\  & A$3.0\substack{+1.5\\ -0.9}$\ & A8.0\\
& $1.4\substack{+0.6 \\ -0.4}\times10^{47}$\ & $4.6\substack{+0.1 \\ -0.2}\times10^{45}$\   &  &  &   &  &\\
\hline
aug1900 & $4.3\substack{+0.3\\ -0.2}$\                                      &  $10.0\substack{+0.04\\ -0.2}$\   & 0.991$\pm$0.001 & $9.3\substack{+2 \\ -2}\times10^{9}$\  & $2.9\substack{+0.9\\ -0.4}\times10^{28}$\  & A$3.1\substack{+1.4\\ -1.1}$\ & A3.9 \\
& $1.5\substack{+0.2 \\ -0.4}\times10^{47}$\  & $3.9\substack{+0.6 \\ -0.1}\times10^{45}$\   & &   & &  &\\
\hline
may1747 & $4.7\substack{+0.05\\ -0.3}$\                     & $10.1\substack{+0.03\\ -0.03}$\                            &  0.984$\pm$0.01  & $4.6\substack{+1 \\ -1}\times10^{9}$\ & $2.1\substack{+0.5\\ -0.5}\times10^{28}$\   & A$1.5\substack{+0.2\\ -0.01}$\ & A4.3 \\
  & $5.0\substack{+0.1 \\ -0.7}\times10^{46}$\  & $1.2\substack{+0.1 \\ -0.03}\times10^{45}$\  &   & &  &  &\\
\hline
aug1909 & $4.0\substack{+0.1\\ -0.6}$\                    & $8.2\substack{+0.3\\ -0.1}$\                              & None  & $1.8\substack{+1 \\ -0.3}\times10^{10}$\  &  $2.2\substack{+1\\ -0.7}\times10^{28}$\  &  A$3.8\substack{+4.5\\ -2.3}$\ &  \textless A1 \\
& $2.5\substack{+2.9 \\ -0.6}\times10^{47}$\   & $2.6\substack{+0.5 \\ -0.9}\times10^{45}$\  &   & &  &  &\\
\hline
 may1736 & $4.2\substack{+0.1\\ -0.03}$\                   & $9.9\substack{+0.3\\ -1.1}$\                &  0.984$\pm$0.003 & $4.9\substack{+3 \\ -2}\times10^{9}$\  & $2.5\substack{+1\\ -1}\times10^{28}$\  & A$1.2\substack{+0.2\\ -0.3}$\  & A2.8 \\
 & $6.9\substack{+0.8 \\ -1.3}\times10^{46}$\   & $2.8\substack{+1.8 \\ -0.08}\times10^{44}$\  & &  &   &  &\\
\hline
may1940 & $4.0\substack{+0.05 \\ -0.4}$\                     & $7.9\substack{+0.06 \\ -0.8}$\                                  & None & $5.5\substack{+2 \\ -0.5}\times10^{9}$\  & $2.8\substack{+1\\ -0.5}\times10^{28}$\  & A$1.5\substack{+1.0\\ -0.6}$\   & A1.8 \\
 & $9.4\substack{+6.1 \\ -0.8}\times10^{46}$\   &  $6.0\substack{+4.0 \\ -0.8}\times10^{44}$\   &  & &   &  &\\
\hline
may1646 & $3.4\substack{+0.1 \\ -0.07}$\                       & $8.0\substack{+0.09 \\ -0.4}$\                                 &  None & $1.5\substack{+0.2 \\ -0.3}\times10^{10}$\  &  $2.0\substack{+0.3\\ -0.4}\times10^{28}$\  & A$1.6\substack{+0.6\\ -0.5}$\  & A1.6 \\
& $2.1\substack{+0.5 \\ -0.5}\times10^{47}$\  &  $5.0\substack{+1.4 \\ -0.6}\times10^{44}$\   &   & &  &  &\\
\hline
may1606 & $3.7\substack{+0.3 \\ -0.2}$\         & $8.0\substack{+0.1 \\ -0.5}$\   &  None & $1.1\substack{+0.3 \\ -0.1}\times10^{10}$\ & $2.1\substack{+0.8\\ -0.3}\times10^{28}$\    & A$1.7\substack{+1.6\\ -0.4}$\  & A2.3 \\ 
& $1.5\substack{+0.7\\ -0.05}\times10^{47}$\ & $6.1\substack{+2.6\\ -0.9}\times10^{44}$\  &   &  & &  &\\
\hline
may1917* & $3.2\substack{+0.04 \\ -0.03}$\                    & $6.4\substack{+0.3 \\ -0.1}$\                                & None  &  &  & A$1.8\substack{+0.3\\ -0.2}$\ & \textless A1\\
& $3.0\substack{+0.4\\ -0.4}\times10^{47}$\   & $6.1\substack{+2.4\\ -2.5}\times10^{44}$\   &   &  & &  &\\
\enddata
\tablecomments{ For aug1850, the thermal parameters from the  \texttt{vapec} component of the \texttt{vapec+bknpwr} fit were used to calculate thermal energy and loop density. For may1917, no obvious thermal volume was seen in a flare-time differenced AIA Fe XVIII image, so density and energy estimates have been omitted.}
\end{deluxetable*}

\begin{deluxetable*}{c|c|ccc|ccc}
\tabletypesize{\footnotesize}
\tablecaption{Event Parameters ( \texttt{vapec+bknpwr} ) \label{nontherm}} 
\tablehead{ \textbf{Event} &  \textbf{VAPEC} &   \textbf{BKNPWR} & & & \colhead{\textbf{GAIN SHIFT}} &  \colhead{\textbf{Energy}} & \colhead{\textbf{GOES}} \\
  &  T (MK) &   PhoIndx2 &  Break Energy & norm  & \colhead{} &  \colhead{Nonthermal} & \colhead{obs.} \\
    &  EM (cm\textsuperscript{-3})  &   &  (keV) & [$ \frac{photons}{keV cm^{2} s} $\  at 1 keV]  & \colhead{} &  \colhead{(erg)} & \colhead{} 
  }
\startdata
aug1850 & $10.3\substack{+0.2\\ -0.2}$\                                                   & $6.3\substack{+0.6\\ -0.5}$\ & $6.0\substack{+0.2\\ -0.3}$\  & $301\substack{+34\\ -33}$\  & 0.972$\pm$0.007 (A) & $7.6\substack{+3\\ -2}\times10^{29}$\ & A7.7 \\
               & $5.6\substack{+0.3 \\ -0.3}\times10^{45}$\ &                                              && & 0.979$\pm$0.007 (B)  &    &\\
\hline
aug1918 & $6.4\substack{+0.04\\ -0.04}$\                                                    & $10.0\substack{+0.6\\ -0.5}$\  & $6.4\substack{+0.1\\ -0.1}$\  & $93\substack{+8\\ -7}$\ & 0.984$\pm$0.003 &  $7.7\substack{+1\\ -1}\times10^{29}$\   &  A4.5 \\
               &  $4.4\substack{+0.2 \\ -0.2}\times10^{47}$\ & &  &   &    &\\
\hline
aug1900 & $6.5\substack{+0.1\\ -0.1}$\                                                            & $10.3\substack{+0.7\\ -0.6}$\  & $6.4\substack{+0.1\\ -0.1}$\  & $320\substack{+27\\ -28}$\ & 0.991$\pm$0.001 & $4.3\substack{+0.7\\ -0.5}\times10^{29}$\  & A3.9\\
               & $3.5\substack{+0.2 \\ -0.2}\times10^{46}$\  &&  &      &  &\\
\hline
\enddata
\end{deluxetable*}

Spectroscopy of the \textit{NuSTAR} microflares was performed using the XSPEC spectral fitting software \citep{1996ASPC..101...17A}. Data from \textit{NuSTAR}'s two telescopes and their corresponding focal plane modules (FPMA, FPMB) were fit simultaneously using the Cash statistic (CSTAT) as a fit statistic, which better handles low-count data \citep{1979ApJ...228..939C}. \par

The same stable, single-CHU-combination time intervals described in Section \ref{space} were used for spectroscopy in 10/11 microflares. As previously noted, \textit{NuSTAR} pointing was reconstructed using several different CHU combinations during event aug1850. However, large pointing shifts were not associated with the CHU changes during the first three minutes (rise/peak) of the microflare. Because of this, aug1850 spectroscopy was performed using data from all component CHU combinations over that interval. \par

Though the \textit{NuSTAR} energy range is typically cited as 3-79 keV, the instrument is capable of observation down to 2 keV. However, differing pixel thresholds complicate the response at the very low end of the energy range, and spectroscopy below 2.5 keV is not recommended \citep{Grefenstette_2016}. As a conservative approach, a lower energy bound of 3 keV was employed for spectroscopy in this analysis. The upper bound of the fit energy range was set to be 10 keV (for fainter events with little to no emission above that energy) or 12 keV for the brightest events (aug1850, aug1900, aug1918, may1618, may1747). The \textit{NuSTAR} spectra were binned to have a minimum of 10 counts in each energy bin. \par 

For each flaring interval, the effects of pileup were estimated by examining the incidence of multiple-pixel events with geometries that cannot be explained by charge-sharing between pixels for a single photon (see Appendix C of \cite{Grefenstette_2016}). Pile-up was found to be negligible in all events. Additionally, it is noted that when a pile-up correction was performed for the brightest microflare considered here (aug1850) in an earlier study \citep{Glesener_2020}, it resulted in changes to spectral parameters small enough to be consistent within their uncertainties (Glesener, personal communication). Because of this, no pile-up correction was performed for the microflares examined here.\par

While the events considered in this study are small in magnitude compared to the population of flares observed by \textit{RHESSI} and other previous solar observatories, they also include the brightest flares yet observed with \textit{NuSTAR} under optimal observing conditions. Because of this, analysis of these events led to the first identification of variations in the \textit{NuSTAR} gain in the extremely low-livetime (\textless 1\%) regime. \par

This phenomenon is described in detail in Appendix \ref{gain}, the conclusion of which is a simple correction to the slope of the linear gain, as well as a set of criteria for determining whether such a correction is likely to be necessary for a given event. Consideration of such corrections will be standard practice for future \textit{NuSTAR} microflare studies. Here, gain corrections were performed for aug1850, aug1900, aug1918, may1618, may1736, and may1747 with the percent shift in gain slope recorded in Table \ref{therm}. In most cases, the gain slope parameter is tied between the two FPM, with uncertainty in the resulting value defined as the difference between FPMA and FPMB gain slope values when the fit is re-run with the FPMA, FPMB gain slopes untied. The orientation of the detector chip gap in FPMA over a portion of the flare site during aug1850 led to significant differences in flux between the two FPM; because of this, the FPMA and FPMB gain slope corrections were determined independently for that event. \par

Initial spectroscopy used XSPEC's isothermal \texttt{vapec} model, which allows for user-specified abundances (taken from \citet{Feldman_1992}, as is standard practice for \textit{NuSTAR} solar spectroscopy \citep[e.g.][]{Wright_2017}). High energy excess was seen over the single \texttt{vapec} model in all eleven events. The origin of this excess was hypothesized to be either emission from smaller volumes of higher temperature plasma likely produced at or near the reconnection site, or nonthermal emission from flare-accelerated electrons. To investigate these possibilities, a second isothermal model was added (\texttt{vapec+vapec}), and a separate fit was also performed using an isothermal model in conjunction with a nonthermal broken power law (\texttt{vapec+bknpwr}). For the \texttt{bknpwr} model component, the spectral index below the break energy was fixed to 2 for all events, as it was expected that the thermal component would dominate at lower energies. \par

\subsection{Spectral Results}\label{specrez}

The majority (8/11) of the microflares were best fit by the \texttt{vapec+vapec} model, as determined both by the use of the CSTAT, as well as manual inspection of residuals. This included all May 2018 microflares and aug1909. Example spectra from one of these clearly thermal microflares are shown in Figure \ref{fig:spec_ex}, displaying all three potential models. The thermal parameters found for these events are reported in Table \ref{therm}, which also reports the estimated thermal energy present in each (the energy values reported are the sum of the thermal energies of both component thermal models). Thermal energies of each component were calculated assuming an isothermal plasma volume with energy given by, 

\begin{equation}
U_{T} = 3k_{B}T \sqrt{EMfV}
\tagaddtext{[erg]}
\end{equation}

\noindent
where T is the temperature of the plasma, EM the emission measure, V the volume, and $f$ is a filling factor (assumed here to be unity). Differenced AIA Fe XVIII images (see Section \ref{space}) were used to estimate a volume for each event by considering the geometries of the Fe XVIII loops deemed most likely to be associated with the \textit{NuSTAR} emission, and converting from the resulting area, A, to a volume (by taking A\textsuperscript{3/2}). These volumes were additionally used to calculate the density of the thermal plasma in each microflare. Densities are also reported in Table \ref{therm}, along with the estimated background-subtracted observed \textit{GOES} class, and the expected GOES class calculated from the \textit{NuSTAR} T and EM using the \texttt{goes\_flux49.pro} IDL routine. The \texttt{goes\_flux49.pro} routine calls CHIANTI version 7.1, and was set to assume coronal abundances \citep{1997A&AS_dere, Landi_2013}.
\par

For two more microflares (aug1900, aug1918), the \texttt{vapec+vapec} models were still unable to fully account for some of the highest energy emission (\textgreater 8 keV - see Figure \ref{fig:spec_ex} for spectra from aug1900). For both of these events, the \texttt{vapec+bknpwr} models did slightly better at higher energies, but found the break energy of the broken power law distribution to occur between 6-7 keV, near the strong Fe complex centered around 6.7 keV. This weakened the case for the \texttt{vapec+bknpwr} model, because at least some of its success could be attributed to the benefit of a break in the spectrum placed near where the thermal continuum is broken by the presence of that Fe complex. \par

For these two events, it cannot be definitively shown that a nonthermal component is present. Perhaps the high-energy excess above the \texttt{vapec+vapec} fit is the result of a multi-thermal plasma more complex than that which can be well-represented by only two isothermal models. To characterize the flare plasma suggested by the thermal interpretation, the \texttt{vapec+vapec} parameters and thermal energies for these events are reported in Table \ref{therm}. \par

For the last event (aug1850), the \texttt{vapec+vapec} fit failed to arrive at a result involving physically realistic plasma temperatures (the higher-temperature found was $\sim$4$\times10^{9}$\ K). In this case, the observed \textit{NuSTAR} spectrum was clearly best fit by the \texttt{vapec+bknpwr} model, as was also found in \cite{Glesener_2020}. The energy content in nonthermal electrons was calculated by determining the nonthermal power from the \texttt{bknpwr} model parameters (assuming a thick target model, as described in \cite{Brown_1971}), and then integrating over the microflare rise times. \par
The resulting nonthermal energy is reported in Table \ref{nontherm}, along with the \texttt{vapec+bknpwr} model parameters and observed/calculated \textit{GOES} classes. The \texttt{vapec} component of the  \texttt{vapec+bknpwr} model is also shown in Table \ref{therm}, where it has been used to estimate a thermal energy content and loop density. The resulting thermal and nonthermal parameters, energies, and density were seen to be qualitatively similar to the results of \cite{Glesener_2020}, though not entirely consistent within uncertainties. The inconsistency is attributed to a slightly different energy range for spectral fitting (3-12 keV here, versus 2.5-12.9 keV) and a modified gain correction procedure (see Appendix \ref{gain}). \par


\section{Discussion}\label{discuss}

\subsection{Assessment of `Large Flare' Properties in \textit{NuSTAR} Microflares}

In this section, these \textit{NuSTAR} microflares are considered with respect to four `large flare' properties: (1) impulsivity in all HXR energies, (2) earlier peak times in higher-energy emission, (3) greater impulsivity in higher-energy emission, and (4) spatial complexity. The temporal evolution of these microflares show generally good agreement with the first two of these properties. As shown in Section \ref{resz}, the majority of the events (8/11) displayed an impulsive time profile in all \textit{NuSTAR} energy ranges considered, and all events were either impulsive or at least consistent with an impulsive profile within uncertainty in the 2-4, 4-6, and 6-8 keV energy ranges. Additionally, 10/11 events (all but may1917, the faintest microflare) display a trend toward earlier peak times in higher-energy \textit{NuSTAR} emission. Both of these results provide evidence that the same energy release processes that drive much brighter events may also be at work at the microflare scale. \par 

The third `large flare' property– greater impulsivity in higher-energy emission– was not observed. However, comparison of impulsivity across energy ranges is complicated by \textit{NuSTAR}'s livetime limitations. Extemely low livetime at microflaring times (combined with proportionally much greater observed flux in the lower end of the \textit{NuSTAR} energy range) limits the spectral dynamic range \citep{Grefenstette_2016}, meaning that higher energy time profiles have poorer statistics and greater uncertainties in the event asymmetry index (see Table \ref{times2}). Thus, the failure to confirm this relationship is not seen as proof that it does not exist for these events. Examination of this property in HXR flares at this scale likely requires an HXR instrument with sensitivity similar to \textit{NuSTAR} that is optimized for the high flux associated with solar observation. \par

Spatial complexity in HXR emission (the fourth property considered) is a standard feature of larger flares, and has also been observed in some \textit{RHESSI} A- and B- class microflares \citep[e.g.][]{2011SSRv..159..263H}. Differential centroid locations between differing HXR energy ranges could result from either a thermal plasma with a spatial gradient in temperature, or distinct thermal and nonthermal HXR sources. Of these eleven events, there were two microflares for which background subtraction could be performed to isolate flare-specific emission from that of the larger, cooler surrounding active region. These both originated from the same set of loops in AR 12712, and neither displayed differences in the centroids of emission in different energy ranges outside our range of uncertainty. This could imply the observation of emission from co-spatial thermal components, or, in the non-thermal interpretation, a situation in which loop-top flare-accelerated electrons are thermalized before reaching loop footpoints (as was concluded in \cite{Glesener_2020}). 
\begin{figure}
\begin{center}
\includegraphics[width=\linewidth]{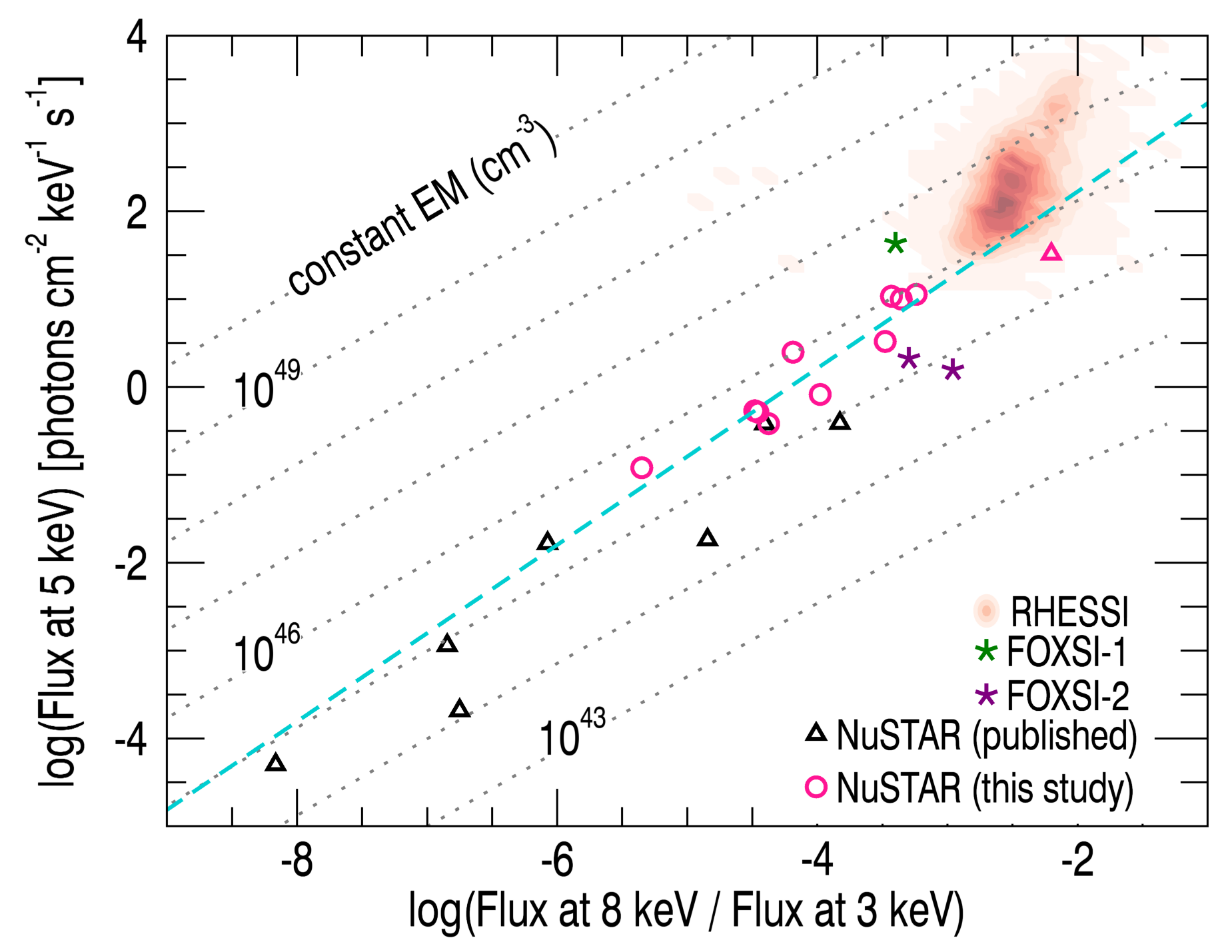}
\caption{ \textit{NuSTAR} microflares from this study (pink circles) are shown in context with other \textit{NuSTAR} microflares (black triangles), as well as \textit{FOXSI} (stars) and \textit{RHESSI} (red) events (aug1850 is included as a pink triangle, to indicate that it is both previously published and a part of this analysis). The vertical axis shows flux at 5 keV (a measure of intensity), while the horizontal shows the ratio of flux at 8 and 3 keV (a temperature analogue), with reference lines corresponding to constant EM. This allows for a comparison of flaring events that is agnostic to any particular spectral model, and shows a strong correlation between these two quantities in the included solar brightenings. A linear fit to the thermal \textit{NuSTAR} microflares from this paper (excluding aug1850) as well as one additional \textit{NuSTAR} thermal flare \citep{Glesener_2017} is shown in blue.}
\label{fig:comp}
\end{center}
\end{figure}
A similar lack of spatial complexity was also seen in the \textit{NuSTAR} microflare examined in \cite{Glesener_2017}, while, contrastingly, \textit{FOXSI} HXR microflare emission was shown to be spatially complex in \cite{2019PhDT........25V}. Further studies involving a greater number of HXR microflares of A-class and below are necessary to determine the relative incidence of these two contrasting results, and also to investigate if they are connected to other microflare properties. \par

\subsection{Thermal vs. Nonthermal Interpretation}

\begin{deluxetable*}{c|cc|c}
\tablecaption{\textit{NuSTAR} Microflare Spectral Models \& Target IDs \label{models}} 
\tablehead{ \colhead{\textbf{Event}} &  \colhead{\textbf{Models}} &   & \colhead{\textbf{Target ID}}}
\startdata
\textbf{\citep{Wright_2017}} & \textbf{VAPEC}   & \textbf{VAPEC}  & SOL2015-04-29T11:29 \\
               & T:   $4.1\substack{+0.2\\ -0.6}$\ MK           &  T:      $10.0\substack{+0.4\\ -1.9}$\ MK      &  \\
               & EM:     $3.1\substack{+3.3 \\ -0.8}\times10^{46}$\   cm\textsuperscript{-3}         &  EM:     $1.4\substack{+3.7 \\ -0.5}\times10^{46}$\   cm\textsuperscript{-3}       &  \\
\hline
\hline
\hline
\textbf{\citep{Glesener_2017}} & \textbf{VAPEC*}  & \textbf{VAPEC} & \cellcolor{red!10} SOL2015-09-01T04:00 \\
               & T:     3.9 MK         &  T:      $6.4\substack{+0.3\\ -0.7}$\ MK      &  \\
               & EM:   $1.0\times10^{46}$\   cm\textsuperscript{-3}       &  EM:    $2.4\substack{+1.49 \\ -0.56}\times10^{45}$\   cm\textsuperscript{-3}        &  \\             
\hline
\hline
\hline
\textbf{\citep{Hannah_2019}} & \textbf{VAPEC*}   & \textbf{VAPEC}  & SOL2016-07-26T23:35 \\
               & T:     $3.23$\ MK         &  T:     $5.08\substack{+0.24\\ -0.66}$\ MK       &  \\
               & EM:     $4.37\times10^{46}$\   cm\textsuperscript{-3}         &  EM:     $6.17\substack{+6.80 \\ -1.99}\times10^{44}$\   cm\textsuperscript{-3}       &  \\
\hline
\hline
\hline
\textbf{\citep{Kuhar_2018}} & \textbf{VAPEC}   &  & \\
Flare 1               &   T:       $3.96\substack{+0.05\\ -0.40}$\ MK       &   &  SOL2016-07-26T21:24\\
               & EM:       $8.5\substack{+6.3 \\ -0.9}\times10^{44}$\   cm\textsuperscript{-3}       &        &  \\
\hline
Flare 2               &   T:       $4.02\substack{+0.05\\ -0.22}$\ MK       &    &  SOL2017-03-21T19:04\\
               & EM:       $1.28\substack{+0.44 \\ -0.16}\times10^{44}$\   cm\textsuperscript{-3}       &      &  \\
\hline
Flare 3               &   T:       $3.28\substack{+0.13\\ -0.06}$\ MK       &    &  SOL2017-03-21T19:30\\
               & EM:     $5.3\substack{+1.8 \\ -1.8}\times10^{44}$\   cm\textsuperscript{-3}         &      &  \\
\hline
\hline
\hline
\textbf{\citep{Glesener_2020}} & \textbf{VAPEC}  & \textbf{BKNPWR} &  \\
aug1850 &  (Table \ref{nontherm})  &  (Table \ref{nontherm}) &  \cellcolor{red!10} SOL2017-08-21T18:50  \\
\hline
\hline
\hline
\textbf{This Work:} & \textbf{VAPEC}  & \textbf{VAPEC}  &  \\
aug1900 & (Table \ref{therm}) &  (Table \ref{therm}) &  \cellcolor{red!10} SOL2017-08-21T19:00 \\
\hline
aug1909 & (Table \ref{therm}) &  (Table \ref{therm}) &  \cellcolor{red!10} SOL2017-08-21T19:09 \\
\hline
aug1918 & (Table \ref{therm}) &  (Table \ref{therm})& \cellcolor{red!10} SOL2017-08-21T19:18  \\
\hline
may1606 & (Table \ref{therm}) &  (Table \ref{therm}) & \cellcolor{red!10} SOL2018-05-29T16:06 \\
\hline
may1618 & (Table \ref{therm}) &  (Table \ref{therm})& \cellcolor{red!10} SOL2018-05-29T16:18  \\
\hline
may1646 & (Table \ref{therm}) &  (Table \ref{therm}) & \cellcolor{red!10} SOL2018-05-29T16:46 \\
\hline
may1736 & (Table \ref{therm}) &  (Table \ref{therm})& \cellcolor{red!10} SOL2018-05-29T17:36 \\
\hline
may1747 & (Table \ref{therm}) &  (Table \ref{therm}) & \cellcolor{red!10} SOL2018-05-29T17:47 \\
\hline
may1917 & (Table \ref{therm}) &  (Table \ref{therm}) & \cellcolor{red!10} SOL2018-05-29T19:17 \\
\hline
may1940 & (Table \ref{therm}) &  (Table \ref{therm}) & \cellcolor{red!10} SOL2018-05-29T19:40 \\
\hline
\hline
\hline
\textbf{\citep{Cooper_2020}} & \textbf{VAPEC*}   & \textbf{VAPEC}  & SOL2018-09-09T11:04 \\
               & T:      $3.20$\ MK        &  T:     $6.66\substack{+0.69\\ -0.71}$\ MK       &  \\
               & EM:      $1.74\times10^{46}$\   cm\textsuperscript{-3}        &  EM:      $0.80\substack{+0.67 \\ -0.32}\times10^{44}$\   cm\textsuperscript{-3}      &  \\
\hline
\enddata
\tablecomments{ Events with target IDs shaded pink were included in the fit line in Figure \ref{fig:comp}. Thermal models with parameters reported without uncertainty (*) were fixed thermal background components. }
\end{deluxetable*}

As described in Section \ref{specrez}, the majority (8/11) of the microflares were found to be best fit by a double thermal model, and the brightest event (aug1850) was found to be best fit by a single thermal model combined with a nonthermal broken power law distribution. The other two (aug1900, aug1918) were similarly well-fit by both double thermal and thermal + broken power law models. \par

To further explore a nonthermal interpretation in those two events, nonthermal energies were calculated from their best fit broken power law model components, assuming a thick target model \citep{Brown_1971}. These energies are reported in Table \ref{nontherm} along with the thermal and broken power law model parameters. The nonthermal energies found are around an order of magnitude larger than derived thermal energies (Table \ref{therm}). This suggests that, while there is not sufficient spectral evidence to prove the presence of a nonthermal electron distribution in these events, the spectra are consistent with a nonthermal source that could power the observed thermal emission. \par

Figure \ref{fig:comp} shows a brightness vs. hardness diagram, which includes these eleven microflares in context with previous \textit{NuSTAR}, \textit{FOXSI} and \textit{RHESSI} events. This representation displays trends in HXR spectral shape across the flares, regardless of their multithermal or nonthermal natures. As discussed in Section \ref{hardrad}, spectral hardness (defined here as the ratio of fluxes at two HXR energies in the continuum) provides a measure of temperature if the flares are isothermal. This parametrization allows for inclusion of multithermal and nonthermal flares in the visualization.  The vertical axis is a measure of intensity at an energy covered by all the instruments in question. For reference, lines of constant EM for an isothermal plasma are overplotted. \par

Additionally, a blue line shows a linear fit to all of the thermal \textit{NuSTAR} flares with significant counts at 8 keV (excluding the nonthermal aug1850). These are the brighter events observed; see spectral models and references for all published \textit{NuSTAR} microflares in Table \ref{models}, where events included in this fit are highlighted. Despite not being included in the fit calculation, aug1850 lies close to this line, confirming that its spectral shape is not at all unusual when compared to the other flares. This hints that the smaller flares may also have nonthermal aspects that are more challenging to disentangle. In particular, aug1900 is noted as a particularly compelling suspect for a hidden nonthermal component, as it not only showed ambiguity between nonthermal and thermal spectral models, but also occurred immediately after aug1850, and from the same set of flare loops within the larger active region (see Figure \ref{fig:img}; Figure 2 in \cite{Glesener_2020}). \par

\section{Conclusions}

In this paper, we have considered eleven \textit{NuSTAR} microflares, ten of which are new to the literature, in order to broaden understanding of the properties of HXR flares at this scale. Consideration of these events together with previous studies \citep{Glesener_2017, Wright_2017, Hannah_2019} begins to establish a picture of a `standard' low-A-class HXR microflare. These events commonly display impulsive time profiles, with higher HXR energies peaking before lower energy HXRs (and before peaks in lower-energy instruments). Their spectra are dominantly thermal, with flare plasma distributions well-approximated by a combination of a brighter, cooler, plasma volume (T=3-5 MK) with a fainter, hotter one (T=5-10 MK). \par

While the presence of nonthermal emission cannot be definitively established in the majority of cases (the nonthermal behavior of the brightest event (aug1850) remains a singular occurrence among microflares observed so far), the spectra of some of the larger events (aug1900, aug1918) are consistent with a picture involving a nonthermal energy source. Therefore, it seems that the range of magnitudes in peak HXR flux spanned by the microflares observed by \textit{NuSTAR} so far includes the transition between a regime where nonthermal emission is dominant, and one where it is largely indistinguishable from thermal emission. Further exploration of nonthermal properties in HXR events of similar brightness is needed to characterize this transition, which is noted as an especially crucial regime for developing an understanding of particle acceleration at the smallest scales. Such exploration will begin with future \textit{NuSTAR} microflare observations, but will require a solar-dedicated focusing HXR instrument to be approached in a statistical manner. \par

\acknowledgments

This work was supported under a NSF Faculty Development Grant (AGS-1429512) to the University of Minnesota, the 2019 NASA Fellowship Program (80NSSC19k1687), an NSF CAREER award (NSF-AGS-1752268), the SolFER DRIVE center (80NSSC20K0627) and the NASA \textit{NuSTAR} Guest Observer program (80NSSC18K1744). The authors also wish to thank Pascal Saint-Hilaire for help with eclipse predictions in planning the 2017 August 21 \textit{NuSTAR} observation.


\appendix

\section{Gain Corrections}\label{gain}

\begin{figure}
\begin{center}
\includegraphics[width=0.6\textwidth]{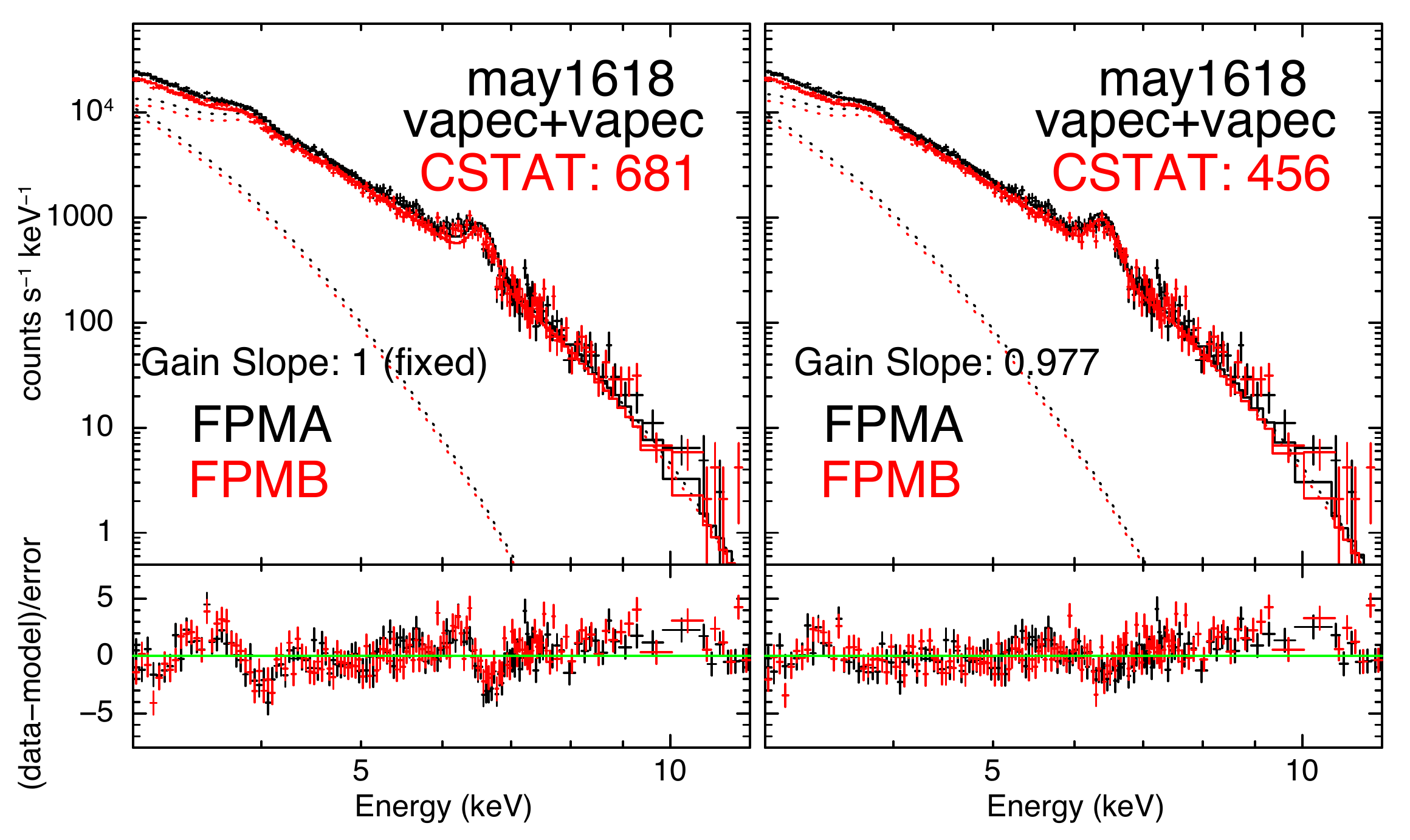}
\caption{Two double thermal (\texttt{vapec+vapec}) fits are shown to the spectrum of microflare may1618 (16:18–16:24 UTC). Livetime-corrected FPMA (black) and FPMB (red) count spectra are shown in each top panel, while the lower panels show the error-normalized residuals. In the left example, the \texttt{vapec+vapec} fit struggles to account for line features between 3-4 and 6-7 keV (clearly seen in residuals). In the right example, the \texttt{vapec+vapec} fit has been repeated with the gain slope parameter freed. Significant improvement is seen in the handling of both emission lines with a small change in gain slope (\textless 3\%). }
\label{fig:gain_ex}
\end{center}
\end{figure}

For several of the brighter microflares (may1618, may1736, may1747, aug1900, aug1918), initial spectroscopy using \texttt{vapec} models resulted in features in fit residuals that indicated a systematic failure of these models to accurately fit the \textit{NuSTAR} spectrum. Specifically, double thermal \texttt{vapec+vapec} models struggled to accurately locate two emission line features present above the thermal continuum (see Figure \ref{fig:gain_ex}, Left). \par

From the CHIANTI atomic database, we expect lines in this energy range observed from hot flaring plasma to be a Ca line at $\sim$3.9 keV (Ca XIX), as well as a complex of lines from transitions in highly-ionized Fe centered around 6.7 keV (the Fe XXV resonance line, along with a collection of Fe XXIV satellites) \citep{1997A&AS_dere, 2015A&A...582A..56D}. When supplied with coronal abundances, the \texttt{vapec} models should be able to accurately represent these features, which are well-understood components of emission from solar plasma at coronal temperatures. \par 

The entire catalogue of observed \textit{NuSTAR} microflares were re-examined in XSPEC, using a constructed fit designed to examine the handling of these emission lines: a sum of two continuum-only thermal models (\texttt{nlapec}) with two fixed-width gaussians to simulate line features. The expected 6.7 keV Fe complex was not found in any \textit{NuSTAR} microflare with sufficient higher-energy statistics to well-locate a line in that energy range; in every case, the line feature observed above 6 keV was found to be shifted lower in energy (often to around 6.4 keV, a difference far greater than the stated $\sim$40 eV systematic uncertainty in the \textit{NuSTAR} gain \citep{2015ApJS..220....8M}). \par

This includes several events for which simultaneous co-spatial emission in the AIA 131 \AA\ channel indicates the presence of plasma at temperatures expected to produce the Fe complex. Even in an accidentally-observed decaying X-class flare (observed at a \textit{GOES} $\sim$C-class level by \textit{NuSTAR}, \citep{Grefenstette_2016}, and well-observed by \textit{RHESSI}), the higher energy \textit{NuSTAR} emission is found to be inconsistent with a 6.7 keV line. \par

In order to explain these changes in line energy as the result of doppler shifts, the emitting thermal plasma would need to be traveling \textit{toward} the Sun at a speed ($\sim$10,000 km s\textsuperscript{-1}) that is over an order of magnitude larger (and in the wrong direction) than plasma velocities observed in either coronal mass ejections or upward-moving plasma volumes associated with chromospheric evaporation  \citep{Gosling_1976, 1984ApJ...287..917A}. \par

The possibility of the actual observation of an emission line around 6.4 keV was explored, as 6.39–6.4 keV Fe K$\alpha$ emission has been historically observed in M- and X-class flares by high-resolution spectrometers \citep[interpreted as flare-driven collisional- or photo-ionization of neutral Fe in the photosphere,][]{1986SoPh..103...89E}. This feature has never been identified in an event anywhere close to as faint as these A-class microflares, but the limited spectral resolution \citep[$\sim$1 keV,][]{2002SoPh..210....3L} of \textit{RHESSI} and limited sensitivity of other instruments means that such an observation has likely never previously been possible. However, the Fe K$\alpha$ explanation for the unexpected incidence of a line at 6.4 keV does not resolve the issue of the failure to observe the 6.7 keV complex even in flares where AIA and \textit{RHESSI} context imply it should be observed. Additionally, microflare-driven photospheric Fe ionization would require a photon or nonthermal electron flux that we do not observe. \par

An investigation was conducted to determine if this discrepancy could be resolved by adding a correction to the \textit{NuSTAR} gain. XSPEC allows for fitting of response parameters, specifically the fitting of slope and intercept (offset) parameters describing a linear representation of the gain. Response fitting was performed using standard \texttt{vapec+vapec} models, where both response and model parameters were allowed to vary simultaneously. \par

With variation of under 5\% in the value of the gain slope only (with no change to the offset), the \texttt{vapec+vapec} model was able to achieve a dramatically better fit to the \textit{NuSTAR} spectrum and resolve the line location discrepancy in every event where it was seen. An example is shown in Figure \ref{fig:gain_ex}, where the addition of only one more parameter (freed gain slope) to the fit allows dramatic improvement in the handling of \textit{both} observed line features. This event is representative of results in all cases where the line location discrepancy was identified. \par

The efficacy of this correction across multiple events has led to the conclusion that a small artificial shift in the \textit{NuSTAR} gain is the most likely explanation for the consistently identified discrepancies in solar spectral lines. A 5\% gain shift is inconsistent with observations taken in ``standard" astrophysical observations where the sources produce moderate count rates (\textless 1000 cps) and high livetime. See, for example, the joint observation of the neutral Fe K$\alpha$ complex in Cen A \citep{F_rst_2016}. This would also result in a $\sim$4 keV shift of the 86.54 keV 155-Eu calibration line, which is not observed \citep{2015_gain_memo}. We therefore conclude that the extreme count rates (\textgreater $10^{5}$\ cps) and the resulting low-livetimes (\textless 1\%) present in many observations of solar active regions and microflares result in a reduced gain of the readout electronics. Such an effect has previously been suggested for the NuSTAR detectors \citep{2012PhDT.........5B}, but until now has not been observed in any astrophysical sources.  \par

The nature of this problem limits the circumstances in which it can be definitively identified: without any significant line features, the continuum thermal spectrum of a plasma volume with a small gain shift added is indistinguishable from the spectrum of a plasma volume possessing slightly different temperature and emission measure. It is only when a line is noticeable above the continuum that a gain discrepancy can be readily observed and quantified by its displacement. For \textit{NuSTAR} solar observations so far, this has so far occurred only in a livetime regime of $\sim$1\% or below. \par

The following procedure is prescribed for investigating possible gain shifts when considering low-livetime \textit{NuSTAR} solar specta, and correcting for them if they are identified. It will be considered a standard aspect of \textit{NuSTAR} solar spectroscopy moving forward:
\begin{itemize}

  \item For events with a line between 6-7 keV that can be located to an energy lower than the expected 6.7 keV, a gain correction should be found by performing a standard \texttt{vapec+vapec} fit with the gain slope parameter freed. 
  
  \item The resulting correction to the gain slope should be applied as a fixed correction for spectral fitting with other model combinations. This is recommended even when \texttt{vapec+vapec} does not give the best fit, as a line location discrepancy can be most accurately identified and corrected for when assuming thermal models only.  
  
  \item For events with no noticeable line features (or, in the event that a visible line is present with no discrepancy between its expected and observed location), the application of a gain correction is not recommended. If a gain correction is applied, it should be understood that improvement in fit does not, on its own, imply the necessity of a correction. Uncertainty ranges for fit parameters should be therefore be extended to include their values when no correction is applied. 
  
\end{itemize}

In events where a correction is deemed necessary (six of the microflares presented here, as well as other yet-unpublished events), the application of gain corrections according to this method has not been seen to have a dramatic effect on fit parameters. For the six events in this paper, the largest changes in the best fit parameters were 10\% in temperature and 40\% in emission measure\footnote{Changes in aug1850 \texttt{bknpwr} parameters– Break Energy: 3\%, Photon Index 2: 10\%, norm: 2\%}. All \textit{NuSTAR} microflare studies published prior to the identification of this issue have been examined to see if gain corrections should be applied retroactively, and none were found to fit the criteria established here. Therefore, it is not expected that any possible gain discrepancy would have affected those scientific results. \par

It is noted that the previous paper considering microflare aug1850 \citep{Glesener_2020} was completed at a time when the gain discrepancy had been identified, but before this standardized procedure had been established. As such, the gain correction applied in that work was performed by freeing the gain slope during a \texttt{vapec+bknpwr} fit, rather than by the method described here. The qualitative agreement in spectral results for aug1850 between that paper and this one show that this difference does not affect the earlier conclusions regarding that microflare. \par

\bibliography{microflare_paper}{}
\bibliographystyle{aasjournal}

\end{document}